\documentclass{ptephy}

\usepackage{amsmath,amssymb}
\usepackage{mathdots}

\makeatletter
\@addtoreset{equation}{section}

\makeatother

\newcommand{\be}{\begin{equation}}
\newcommand{\ee}{\end{equation}}
\newcommand{\bea}{\begin{eqnarray}}

\newcommand{\beann}{\begin{eqnarray*}}
\newcommand{\eeann}{\end{eqnarray*}}
\newcommand{\nn}{\nonumber}
\newcommand{\ba}{\begin{array}}
\newcommand{\ea}{\end{array}}

\newcommand{\Z}{\mathbb{Z}}
\newcommand{\R}{\mathbb{R}}
\newcommand{\C}{\mathbb{C}}

\newcommand{\del}{\partial}

\newcommand{\ta}{{\tilde{a}}}

\DeclareMathOperator{\Tr}{Tr}
\newcommand{\cO}{{\mathcal O}}

\begin{document}

\title{
Ising Model on Twisted Lattice and Holographic RG flow
}

\author{
\name{\fname{So} \surname{Matsuura}}{1}, and
\name{\fname{Norisuke} \surname{Sakai}}{1}
}

\address{\affil{1}{Department of Physics, and Research and Education Center 
for Natural Sciences, 
Keio University, Hiyoshi 4-1-1, Yokohama, Kanagawa 223-8521, Japan}
\email{${}^1$s.matsu@phys-h.keio.ac.jp, ${}^2$norisuke.sakai@gmail.com, 
}
}

\begin{abstract}%
The partition function of the two-dimensional Ising model is 
exactly obtained on a lattice with a twisted boundary condition. 
The continuum limit of the model off the critical temperature is 
found to give the mass-deformed Ising conformal field theory 
(CFT) on the torus with the complex structure $\tau$. 
We find that the renormalization group (RG) flow of the mass parameter 
can be holographically described in terms of 
the three-dimensional gravity including a scalar field with
a simple nonlinear kinetic function and a quadratic potential. 
\end{abstract}


\maketitle

\section{Introduction}

The AdS/CFT correspondence\cite{Maldacena:1997re,Gubser:1998bc,Witten:1998qj} 
has been giving valuable informations for various field theories 
especially in the strong coupling regions.
In particular, the correspondence between the classical gravities 
and the large $N$ (gauge) field theories has been most intensively studied. 
Among many tests on this correspondence, 
the exactly soluble models in low dimensions,
say, the two-dimensional conformal field theories, 
are expected to shed light. 
The relation of AdS gravity with the Virasoro algebra of CFT 
is based on the pioneering work of Brown and Henneaux\cite{Brown:1986nw}, 
and the minimal conformal field theories in two dimensions have 
been discussed in relation to the AdS gravity in three 
dimensions\cite{Witten:2007kt,Maloney:2007ud}. 
Subsequently extensive analyses have been done for 
the $W_N$ minimal conformal field theories
and many evidences are presented for the relevance of the higher 
spin field theories together with gravity for these minimal 
models in the large $N$ limit%
\cite{Gaberdiel:2010pz,
Gaberdiel:2012uj,
Henneaux:2010xg,
Campoleoni:2010zq,
Gaberdiel:2011wb,
Campoleoni:2011hg,
Creutzig:2011fe,
Castro:2010ce,
Gaberdiel:2012yb,
Gaberdiel:2012ku,
Perlmutter:2012ds,
Chang:2013izp,
Ferlaino:2013vga,
Gaberdiel:2013jpa,
Fujisawa:2013ima,
Creutzig:2014ula}.

As for the recent progress in the stronger statement 
of the AdS/CFT correspondence, that is, 
the correspondence between 
the ``quantum'' gravity or the string theory 
and the finite $N$ (gauge) field theory, 
the finite $N$ effect of the one-dimensional supersymmetric 
gauge theory with 16 supercharges 
(the finite $N$ BFSS matrix model\cite{Banks:1996vh}) 
has been directly examined 
using the Monte Carlo simulation\cite{Hanada:2008ez,Hanada:2013rga}, 
where the $\alpha'$-corrections to 
the Type IIA supergravity \cite{Hyakutake:2007sm} are reproduced
from the gauge theory. 
This strongly 
motivates to regard field theories in some category as candidates 
of the quantum gravity at least in the semi-classical meaning. 
In fact, in Ref.~\cite{Castro:2011zq}, 
it has been proposed that the two-dimensional minimal models 
are holographically dual to three-dimensional quantum gravities. 
In particular, 
the simplest minimal model, namely the Ising conformal field theory, 
is conjectured to be dual with 
the three-dimensional Euclidean pure quantum gravity.

We here briefly review the argument in Ref.~\cite{Castro:2011zq}. 
Let us consider the three-dimensional quantum gravity with 
negative cosmological constant. 
The ``quantum'' here means that we integrate over 
all the possible three-dimensional metric.
In this path-integral, the boundary of the three-dimensional geometry is
fixed to the 2-torus with the complex structure $\tau$, 
and thus the partition function of the gravity is a function of $\tau$ and $\bar\tau$. 
The important fact is that the smooth classical solutions of the three-dimensional 
pure gravity are restricted to the locally AdS geometries, which are obtained by 
the $SL(2,\Z)$ transformation from the thermal AdS geometry. 
Thus, if we can use the semi-classical approach, the partition function of 
the quantum gravity is obtained by summing up the classical contributions of 
each solution with some quantum corrections. 
The expansion parameter of the saddle point approximation is the inverse 
of the quantity, 
\begin{equation}
c=\frac{3L}{2G_N}, 
\end{equation}
where $L$ is the AdS radius and $G_N$ is the Newton constant. 
The quantity $c$ equals to the central charge of the two-dimensional 
Virasoro algebra which appears at the boundary 
of the asymptotically AdS geometry\cite{Brown:1986nw}.  
Therefore the semi-classical approximation is usually expected 
to work only for large $c$ region. 
The most important assumption in Ref.~\cite{Castro:2011zq} is 
that the partition function can be evaluated by summing up 
the classical geometries even in the strong coupling region, $c\sim 1$. 
Once this assumption is accepted, the partition function can be written as
\begin{equation}
Z_{\rm grav}(\tau,\bar\tau)=\sum_{\gamma\in SL(2,\Z)/\Gamma_c} 
Z_{\rm vac}(\gamma\tau,\gamma\bar\tau), 
\label{sum gamma}
\end{equation}
where $\Gamma_c$ is a subgroup of $SL(2,Z)$ which does not change 
the topology of the three-dimensional geometry 
and $Z_{\rm vac}(\tau,\bar\tau)$ is the contribution from the 
thermal AdS geometry. 
The points are that $Z_{\rm vac}(\tau,\bar\tau)$ can be calculated 
explicitly and that the symmetry $\Gamma_c$ is enhanced at specific 
values of $c<1$ where the summation of (\ref{sum gamma}) becomes 
a finite sum. 
Note that these values exactly equal to the central charges 
allowed for the two-dimensional minimal models. 
In particular, when $c=\frac{1}{2}$, the partition function of the gravity 
(\ref{sum gamma}) reproduces that of the two-dimensional $c=\frac{1}{2}$ 
conformal field theory, that is, the Ising conformal field theory. 
This result suggests the duality between the $c=\frac{1}{2}$ minimal model 
and the three-dimensional quantum pure gravity.

This argument actually gives the correspondence between the $c=\frac{1}{2}$ 
conformal fixed point of the theory space of the two-dimensional field theory 
and the three-dimensional quantum gravity. 
Then it is natural to expect that there will be a gravity description at least 
in the vicinity of the conformal fixed point in the theory space 
of the two-dimensional quantum field theories. 
In the context of the AdS/CFT correspondence, 
the renormalization group (RG) flow of a coupling constant in 
the field theory can be identified with the classical trajectory 
of the corresponding bulk field 
in the asymptotically AdS geometry, which is called the holographic RG 
\cite{Akhmedov:1998vf, Alvarez:1998wr,Freedman:1999gp,
Girardello:1998pd, Girardello:1999bd, Porrati:1999ew, Balasubramanian:1999jd, 
Skenderis:1999mm, DeWolfe:1999cp, de Boer:1999xf, Fukuma:2000bz}
(see also \cite{Fukuma:2002sb}). 
In this scheme, the radial coordinate of the AdS gravity can be identified 
with the RG parameter and the AdS boundary corresponds to the conformal 
fixed point. 
In the case of the Ising model, 
although the conformal fixed point corresponds to the ``quantum'' gravity, 
we can expect that the central assumption in Ref.~\cite{Castro:2011zq}, 
namely, only the classical solutions contribute to 
the partition function of the quantum gravity, 
still works at least in the vicinity of the AdS solution 
even after adding an additional field in the bulk theory. 
Then we can examine the RG structure around the $c=1/2$ conformal fixed point 
using the traditional technique of the holographic RG. 

The purpose of our paper is to work out the exact solution of 
the Ising model partition function on the lattice 
corresponding to the torus with the generic complex structure $\tau$, 
and study the corresponding holographic RG structure of the continuum 
theory around the conformal fixed point. 
Usually the partition function of Ising model in two dimensions 
is obtained for rectangular lattices with the periodic boundary 
condition, 
and one obtains the Ising conformal field theory on a rectangular 
torus ($\tau=i$) by taking the continuum limit at the critical temperature
\cite{onsager,kaufman,ferdinandFisher}. 
However, the most general torus has a complex structure with 
the parameter $\tau$, representing the shape of the torus. 
In the literature, we have found no explicit solution of 
the Ising model partition function on the lattice corresponding 
to the torus with the generic complex structure, although 
there have been works to obtain finitized conformal spectrum 
of Ising model on manifold of various topology using 
corner-transfer matrix, Yang-Baxter technique, or thermodynamic 
Bethe Ansatz\cite{Melzer:1993zk,O'Brien:1996,Chui:2001nu,Feverati:2004bv}. 
We explicitly compute the partition function of the Ising model 
on the twisted lattice, that is, a lattice 
with such a boundary condition 
that the ``space'' position is shifted to some amount 
when one goes around the ``time'' direction, 
and show that it ends up with the torus with the complex structure 
in the continuum limit. 
By taking the continuum limit at off-critical temperature 
with an appropriate scaling, 
we identify the deviation parameter as the mass 
of the Ising conformal field theory on the torus with the 
complex structure. 
We also work out the classical solution of the three-dimensional 
Einstein gravity with a single scalar field. 
We find that a simple nonlinear kinetic function and a simple 
quadratic potential for the scalar field can capture the RG flow from 
the Ising field theory with the central charge $c=1/2$ at the 
ultraviolet fixed point towards the $c=0$ case at the infrared 
by using the technique of the holographic RG 
via the Hamilton-Jacobi equation of the gravity.

This paper is organized as follows:
In the next section, the partition function of the two-dimensional Ising model 
is obtained on the twisted lattice representing the discretized 
version of the torus with the complex structure. 
In section \ref{sc:continuum}, the continuum limit of the partition 
function is obtained retaining the deviation from critical temperature, 
which results in a mass term for a free Majorana fermion. 
In section \ref{sc:holgraphicRGflow}, the holographic description 
of the renormalization group flow is worked out for the massive 
Majorana fermion in terms of the Hamilton-Jacobi equation. 
Section \ref{sc:conclusion_discussion} is devoted to a summary 
of our results and a discussion. 
Some technical details in computing the partition function of the 
Ising model is summarized in Appendix \ref{app:trans}. 
Some details of partition function of two-dimensional massive 
Majorana fermion on the torus is given in Appendix \ref{app:massive fermion}.

\section{Partition function of the 2D Ising model on the twisted lattice}
\label{sc:partition_function}
Let us consider the 2D Ising model on a rectangular lattice 
with the size $n\times m$. 
There is a "spin" $s(x)=\pm 1$ at each site $x=(x_1,x_2)$ 
$(x_1=1,\ldots, n,\, x_2=1,\ldots, m)$ 
and the Hamiltonian of the system is given by
\begin{equation}
H=-J_1 \sum_x s(x)s(x+\hat{1})-J_2 \sum_x s(x)s(x+\hat{2}), 
\label{Ising hamiltonian}
\end{equation}
where $J_1$ ($J_2$) and $\hat{1}$ ($\hat{2}$) are the coupling 
constant and the unit vector in the ``space'' (``time'') direction, 
respectively. 
\begin{figure}[htbp]
\begin{center}
\scalebox{0.4}{
\includegraphics{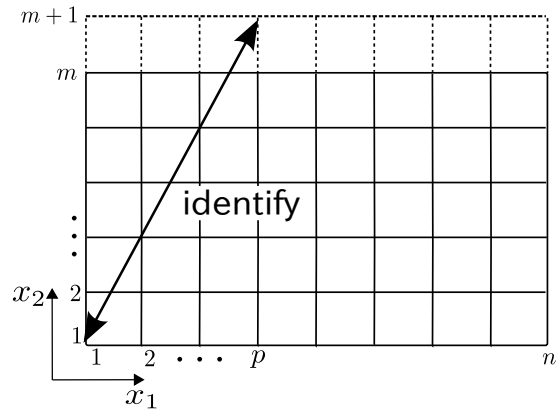}}
\caption{
The spin degrees of freedoms are on the sites of the $n\times m$ lattice. 
In our case, we take the usual periodic boundary condition for the $x_1$ direction 
but we adopt the twisted boundary condition for $x_2$, namely, we identify 
$(x_1+p,x_2+m)$ with $(x_1, x_2)$. 
 }
\label{fig:lattice}
\end{center}
\end{figure}
As for the boundary condition, 
we impose the periodic boundary condition to the space direction 
and the twisted or shifted boundary condition with an integer parameter $p\in\Z$
to the time direction, 
namely, the space position is shifted by $p$ when one goes around the time direction
(see Fig.~\ref{fig:lattice}): 
\begin{equation}
s(x_1+n,x_2) = s(x_1,x_2), \quad 
s(x_1+p,x_2+m) = s(x_1, x_2). 
\label{bdr cond}
\end{equation}
In the following, we evaluate the partition function, 
\begin{equation}
Z_p \equiv \sum_{\{s(x)=\pm 1\}} e^{-\beta H}, 
\label{Ising PF1}
\end{equation}
under this boundary condition.

Let us first define the matrices with the size of $2^n$, 
\begin{equation}
S_k^i \equiv {\mathbf 1}_2 \otimes \cdots \otimes \sigma^i 
\otimes \cdots \otimes \mathbf{1}_2, 
\end{equation}
where the Pauli matrices $ \sigma^i$ ($i=1,2,3$) are 
placed at the $k$-th position ($k=1,\ldots,n$). 
Using these matrices, the transfer matrix of this system is written as 
\begin{equation}
{T} = \left( 2\sinh 2a \right)^{\frac{n}{2}} V_a V_b, 
\end{equation}
with
\begin{align}
V_a \equiv \prod_{k=1}^n \exp\left(\tilde{a} S_k^1\right),  \quad 
V_b \equiv \prod_{k=1}^n \exp\left(b S_k^3 S_{k+1}^3 \right), 
\end{align}
where $a$ and $b$ are given by $a\equiv \beta J_1$ and $b\equiv \beta J_2$
and $\ta$ is defined by 
\begin{equation}
 \sinh 2\ta  \equiv \frac{1}{\sinh 2a } .
 \end{equation}
Note that the system is in the ordered phase for $\tilde{a}<b$ 
and is in the disordered phase for $\tilde{a}>b$. 
We further define the "shift matrix" $\Sigma$ whose components are explicitly 
given by
\begin{equation}
\begin{cases}
\Sigma_{k,2k-1}=\Sigma_{2^{n-1}+k,2k}=1, & (k=1,\cdots,2^{n-1}) \\
0, & ({\rm others})
\label{shift matrix in spin}
\end{cases}
\end{equation}
We see that $\Sigma$ has the property,
\begin{equation}
 \Sigma^{-1} S_k^i \Sigma = S_{k-1}^i, \quad (S_{n+1}^i \equiv S_1^i). 
\end{equation}
We can then write the partition function in Eq.(\ref{Ising PF1}) 
with the boundary condition in Eq.(\ref{bdr cond}) as 
\begin{equation}
Z={\rm Tr} \left( {T}^m \Sigma^p \right). 
\end{equation}

In order to estimate the transfer matrix, 
we here construct the Dirac matrices of $Spin(2n)$ as 
\begin{align}
\Gamma_{2k-1}&\equiv S_1^1S_2^1\cdots S_{k-1}^1 S_k^3, \quad 
\Gamma_{2k} \equiv S_1^1 S_2^1 \cdots S_{k-1}^1 S_k^2,  \quad
(k=1,\ldots, n)
\label{Gamma}
\end{align}
and 
\begin{equation}
U\equiv S_1^1 S_2^1 \cdots S_n^1 = i^n \Gamma_1 \cdots \Gamma_{2n}, 
\end{equation}
which satisfy the Clifford algebra, 
\begin{equation}
\{\Gamma_\mu, \Gamma_\nu\}=2\delta_{\mu,\nu}, \quad 
\{\Gamma_\mu, U\}=0, \quad 
U^2=1. \quad 
(\mu,\nu=1,\cdots,2n)
\end{equation}
As usual, the generators of $Spin(2n)$ in the chiral and anti-chiral representations 
are defined as 
\begin{equation}
 J_{\mu\nu}^{\pm} \equiv \frac{i}{4}[\Gamma_\mu, \Gamma_\nu]\, U_\pm, 
\end{equation}
where 
\begin{equation}
U_\pm \equiv  \frac{1}{2} (1\pm U).
\end{equation} 
Using them, 
we can also divide $V_a$, $V_b$ and $\Sigma$ into the chiral and anti-chiral 
sectors as $V_a^\pm \equiv V_a\, U_\pm$, 
$V_b^\pm \equiv V_b\, U_\pm$ and $\Sigma_\pm \equiv \Sigma\,  U_\pm$. 
In the following, we use the notation, 
\begin{align}
H_\pm &\equiv  \left(V_a^\pm \right)^{1/2} V_b^\pm  \left(V_a^\pm \right)^{1/2}. 
\label{def Hpm}
\end{align}
For $\Sigma_\pm$, we can easily see 
\begin{equation}
\left(V_a^\pm \right)^{-1/2} \Sigma_\pm  \left(V_a^\pm \right)^{1/2} = \Sigma_\pm. 
\end{equation} 
Then the partition function Eq.(\ref{Ising PF1}) can be written as 
\begin{align}
Z =& (2 \sinh 2a )^{\frac{mn}{2}} 
\Tr\left(
 H_+^m \Sigma_+^p U_+ +  H_-^m \Sigma_-^p U_ -
\right) \nn \\
=& (2 \sinh 2a )^{\frac{mn}{2}} \Bigl\{
\Tr_+ \left(  H_+^m \Sigma_+^p \right) 
+ \Tr_- \left( H_-^m  \Sigma_-^p \right) 
\Bigr\},  
\label{Ising PF2}
\end{align}
where $\Tr_\pm$ denote the trace over the chiral and anti-chiral sectors 
of the spin representation of $Spin(2n)$, respectively. 

In evaluating (\ref{Ising PF2}), the following fact is useful: 
Suppose that ${}^{\exists} A\in Spin(2n;\C)$ in the fundamental representation is 
transformed into the ``canonical form'' by $T\in O(2n)$ as 
\begin{equation}
T^T A T = 
\bigoplus_{k=1}^n e^{ \theta_{k} \hat{J}_{2k-1,2k} } 
= \bigoplus_{k=1}^n R(-i\theta_k), \quad (\theta_k \in \C)
\end{equation}
where 
 $\hat{J}_{\mu\nu}$ are the generators of $Spin(2n;{\mathbb C})$ in the 
fundamental representation and 
$R(\theta)$ is the two-dimensional rotation matrix 
with the (complex) angle $\theta$, 
\begin{equation}
R(\theta)\equiv\left(
\begin{matrix}
\cos(\theta) & -\sin(\theta) \\
\sin(\theta) & \cos(\theta)
\end{matrix} \right). 
\label{2d rotation}
\end{equation}
Then we can write $\Tr_\pm (A)$ as
\begin{equation}
{\rm Tr}_\pm (A) = \frac{1}{2} \left( 
\prod_{k=1}^n 2\cosh\frac{\theta_k}{2} 
\pm \det(T)  \prod_{k=1}^n 2\sinh\frac{\theta_k}{2} 
\right). 
\label{trace}
\end{equation}
Thus, we first express $H_{\pm}^m\Sigma_{\pm}^p$ 
in the fundamental representation 
and then transform them into the canonical form 
using appropriate matrices $T_\pm \in O(2n)$.

The computation along this strategy is straightforward and we summarize it 
in Appendix \ref{app:trans}. 
As a preparation to show the result, we define the quantity 
$\gamma_I$ $(I=1,\cdots,2n)$ as the positive solution of the equation, 
\begin{equation}
\cosh \gamma_I = \cosh 2\tilde{a} \cosh 2b - \cos\left(\frac{\pi I}{n}\right) \sinh 2\tilde{a} \sinh 2b, 
\label{gamma-r}
\end{equation}
and 
\begin{equation}
\tilde \gamma_I \equiv \gamma_I - i \frac{\pi p I}{mn}. 
\label{gamma-c}
\end{equation}
Note that $\tilde\gamma_I$ satisfy the relation: 
\begin{equation}
e^{m\tilde\gamma_{2n-I}} = e^{m\tilde\gamma_{I}^{*}}. 
\label{eq:reflection}
\end{equation}
In addition, we will often use $\gamma_0 = \gamma_{2n}$ in the following. 
Combining the above consideration and the results (\ref{canonical1}) and (\ref{canonical2}), 
we obtain the partition function of the two-dimensional Ising model in the ordered phase 
with the twisted boundary condition in Eq.(\ref{bdr cond}); 
\begin{equation}
Z= \frac{1}{2} \left( 2 \sinh 2a \right)^{\frac{mn}{2}} \sum_{i=1}^4 Z_i, 
\label{Ising PF}
\end{equation}
with 
\begin{align}
\begin{split}
Z_1 &= R_1 \prod_{r=1}^{ \left[ \frac{n}{2} \right] } 
\left|  2 \cosh \left( \frac{m}{2} \tilde{\gamma}_{2r-1} \right) \right|^2
\equiv \left(\prod_{k=1}^{n} e^{\frac{m}{2} \gamma_{2k-1} } \right) P_1,  \\
Z_2 &= R_2 \prod_{r=1}^{ \left[ \frac{n}{2} \right] } 
\left|  2 \sinh \left( \frac{m}{2} \tilde{\gamma}_{2r-1} \right) \right|^2
\equiv \left(\prod_{k=1}^{ n } e^{\frac{m}{2} \gamma_{2k-1} } \right) P_2,  \\
Z_3 &= 2 \cosh\left( \frac{m}{2} \gamma_0 \right) R_3 \prod_{r=1}^{ \left[ \frac{n-1}{2} \right] } 
\left|  2 \cosh \left( \frac{m}{2} \tilde{\gamma}_{2r} \right) \right|^2
\equiv \left(\prod_{k=1}^{n} e^{\frac{m}{2} \gamma_{2k} } \right) 
\cdot \left( 1 + e^{-m\gamma_0} \right) P_3,  \\
Z_4 &= 2 \sinh\left( \frac{m}{2} \gamma_0 \right)  R_4 \prod_{r=1}^{ \left[ \frac{n-1}{2} \right] } 
\left|  2 \sinh \left( \frac{m}{2} \tilde{\gamma}_{2r} \right) \right|^2
\equiv \left( \prod_{k=1}^{ n} e^{\frac{m}{2} \gamma_{2r} } \right) 
\cdot \left( 1 - e^{-m\gamma_0} \right) P_4,  \\
\end{split}
\label{func Z}
\end{align}
where we have used the reflection property (\ref{eq:reflection}) 
and introduced 
\begin{align}
\begin{split}
R_1 &= \begin{cases} 
1 & (n:{\rm even}) \\ 
2\cosh\left( \frac{m}{2} \gamma_n \right) & (n:{\rm odd},\ p:{\rm even}) \\
2\sinh\left( \frac{m}{2} \gamma_n \right) & (n:{\rm odd},\ p:{\rm odd}) \\
\end{cases}, \
R_2 = \begin{cases} 
1 & (n:{\rm even}) \\ 
2\sinh\left( \frac{m}{2} \gamma_n \right) & (n:{\rm odd},\ p:{\rm even}) \\
2\cosh\left( \frac{m}{2} \gamma_n \right) & (n:{\rm odd},\ p:{\rm odd}) \\
\end{cases} \\
R_3 &= \begin{cases} 
2\cosh\left( \frac{m}{2} \gamma_n \right) & (n:{\rm even},\ p:{\rm even}) \\
2\sinh\left( \frac{m}{2} \gamma_n \right) & (n:{\rm even},\ p:{\rm odd}) \\
1 & (n:{\rm odd}) 
\end{cases}, \
R_4 = \begin{cases} 
2\sinh\left( \frac{m}{2} \gamma_n \right) & (n:{\rm even},\ p:{\rm even}) \\
2\cosh\left( \frac{m}{2} \gamma_n \right) & (n:{\rm even},\ p:{\rm odd}) \\
1 & (n:{\rm odd})
\end{cases} 
\end{split}
\end{align}
and 
\begin{align}
\begin{split}
P_1 &= \Bigl(  \prod_{r=1}^{\left[\frac{n}{2}\right]} 
\left| 1 + e^{-m\tilde\gamma_{2r-1}} \right|^2  \Bigr)
\Bigl(1+(-1)^{p} \delta_{(-1)^n,-1} e^{-m\gamma_n} \Bigr), \\
P_2 &=  \Bigl(  \prod_{r=1}^{\left[\frac{n}{2}\right]} 
\left| 1 - e^{-m\tilde\gamma_{2r-1}} \right|^2 \Bigr)
\Bigl( 1 - (-1)^p \delta_{(-1)^n,-1} e^{-m\gamma_n} \Bigr) , \\
P_3 &=  \Bigl( \prod_{r=1}^{\left[\frac{n-1}{2}\right]} 
\left| 1 + e^{-m\tilde\gamma_{2r}} \right|^2 \Bigr)
\Bigl( 1 + (-1)^p \delta_{(-1)^n,1}  e^{-m\gamma_n} \Bigr), \\
P_4 &= \Bigl(  \prod_{r=1}^{\left[\frac{n-1}{2}\right]}  
\left| 1 - e^{-m\tilde\gamma_{2r}} \right|^2 \Bigr)
\Bigl( 1 - (-1)^p \delta_{(-1)^n,1}  e^{-m\gamma_n} \Bigr). 
\end{split}
\label{func P}
\end{align}

The result Eq.(\ref{func Z}) constitutes our new result 
of partition function of the two-dimensional Ising model 
on a twisted lattice that gives the discretized version of 
torus with the complex structure $\tau$  
\begin{equation}
\tau=\tau_1+i\tau_2, \qquad 
\tau_1 \equiv \frac{p}{n}, \qquad 
\tau_2 \equiv \frac{m}{n}.
\label{two moduli}
\end{equation}
whose continuum limit gives the torus with the complex 
structure $\tau$ as described in the next section.

\section{Continuum limit of the partition function}
\label{sc:continuum}

We next consider the continuum limit, that is, the limit of 
$m,n,p\to \infty$ with fixing  the ratios $\tau_1, \tau_2$ 
in Eq.(\ref{two moduli}). 
In taking this limit, we also tune the coupling constant so 
that the theory properly reaches to a continuum theory around 
the critical point. 
For simplicity, we consider the case of $J_1=J_2$, that is, 
$a=b\equiv K$. 
Under this condition, we defined the parameter $\mu$ through 
the relation 
\begin{equation}
\frac{\mu^2}{4 n^2} = \frac{1}{2} 
\left( \sinh 2K + \frac{1}{\sinh 2K} \right) - 1. 
\label{eq:deviation_critical_temp}
\end{equation}
Recalling that the critical temperature $K^*$ is given by 
$\sinh 2K^* =1$, the parameter $\mu$ expresses a deviation from 
the critical temperature for a finite $n$ \cite{ferdinandFisher}. 
Note that we take the continuum limit $n\to\infty$ with fixing $\mu={\cal O}(1)$. 
This means that the temperature approaches to the critical value as $K-K^*={\cal O}(n^{-1})$ 
in taking the continuum limit.

In this parametrization, $\gamma_I$ can be expressed as 
\begin{align}
\cosh \gamma_I &= \frac{\mu^2}{2n^2}+1+2\sin^2 \frac{\pi I}{2n}. 
\end{align}
Since we are interested in the continuum limit, only the region $I \ll n$ is relevant. 
Then, in such a region, $\gamma_I$ can be expanded for $n\gg 1$ as 
\begin{align}
\gamma_I&=\frac{2\pi}{n}\sqrt{ \left(\frac{\mu}{2\pi}\right)^2 +  \left(\frac{I}{2}\right) ^2 }
+{\mathcal O}(n^{-3}).
\end{align}
Substituting it to Eq.(\ref{func P}) and taking the limit of $n\to\infty$, 
$P_1,\cdots,P_4$ become
\begin{equation}
\begin{split}
\lim_{n\to \infty} P_1 = 
\prod_{k=0}^{\infty} \left| 
 1+e^{ 2\pi i  \left( k + \frac{1}{2} \right)  \left(  
   \tau_1 
   +i \tau_2 \sqrt{ 1+ \left(\frac{\mu}{(2k+1)\pi} \right)^2}
   \right) }
 \right|^2, \\
 \lim_{n\to \infty} P_2 = \prod_{k=0}^{\infty} \left| 
 1-e^{ 2\pi i  \left( k + \frac{1}{2} \right)  \left(  
   \tau_1 
   +i \tau_2 \sqrt{ 1+ \left(\frac{\mu}{(2k+1)\pi} \right)^2}
   \right) }
 \right|^2, \\
 \lim_{n\to \infty} P_3 = 
 \prod_{k=1}^{\infty} \left| 
 1+e^{ 2\pi i k  \left(  
   \tau_1 
   +i \tau_2 \sqrt{ 1+ \left(\frac{\mu}{2\pi k} \right)^2}
   \right) }
 \right|^2, \\
 \lim_{n\to \infty} P_4 = \prod_{k=1}^{\infty} \left| 
 1-e^{ 2\pi i k  \left(  
   \tau_1 
   +i \tau_2 \sqrt{ 1+ \left(\frac{\mu}{2\pi k} \right)^2}
   \right) }
 \right|^2, 
\end{split}
\label{limit P}
\end{equation}
where we have used the reflection property of $\tilde\gamma_I$ (\ref{eq:reflection}). 
We can also show
\begin{equation}
 \lim_{n\to\infty} e^{-m\gamma_0} = e^{-\tau_2 \mu}, 
 \label{limit pf1}
\end{equation}
and
\begin{align}
 \lim_{n\to\infty} e^{\frac{m}{2}\tilde\gamma_I} 
  = e^{ \pi \tau_2 \sqrt{ \left(\frac{I}{2}\right)^2 
+ \left(\frac{\mu}{2\pi}\right)^2 } 
  -i\frac{\pi}{2} \tau_1 I }. 
  \label{limit pf2}
\end{align}
Combining Eqs.(\ref{limit P}), (\ref{limit pf1}) and (\ref{limit pf2}), 
we see that the partition function in the continuum limit 
can be expressed as 
\begin{equation}
Z={\mathcal C}\sum_{i=1}^4 Z_i^{\rm cont.}, 
\label{cont PF}
\end{equation}
where ${\mathcal C}$ is an irrelevant constant and 
$Z_i^{\rm cont.}\equiv\lim_{n\to \infty}Z_i$ 
are given by 
\begin{equation}
\begin{split}
Z_1^{\rm cont.} &= 
 \left(
 \prod_{k \in {\mathbb Z}} 
 e^{ \pi \tau_2 \sqrt{ \left(k+\frac{1}{2} \right)^2 
+ \left(\frac{\mu}{2\pi}\right)^2 }  }
 \right)
\prod_{k=0}^{\infty} \left| 
 1+e^{ 2\pi i  \left( k + \frac{1}{2} \right)  \left(  
   \tau_1 
   +i \tau_2 \sqrt{ 1+ \left(\frac{\mu}{(2k+1)\pi} \right)^2}
   \right) }
 \right|^2, \\
Z_2^{\rm cont.} &= 
 \left(
 \prod_{k \in {\mathbb Z}} 
 e^{ \pi \tau_2 \sqrt{ \left(k+\frac{1}{2} \right)^2 
+ \left(\frac{\mu}{2\pi}\right)^2 }  }
 \right)
\prod_{k=0}^{\infty} \left| 
 1-e^{ 2\pi i  \left( k + \frac{1}{2} \right)  \left(  
   \tau_1 
   +i \tau_2 \sqrt{ 1+ \left(\frac{\mu}{(2k+1)\pi} \right)^2}
   \right) }
 \right|^2, \\
Z_3^{\rm cont.} &= 
 \left(
 \prod_{k \in {\mathbb Z}} 
 e^{ \pi \tau_2 \sqrt{ k^2 + \left(\frac{\mu}{2\pi}\right)^2 }  }
 \right)
 \left(
 1+e^{-\tau_2 \mu} 
 \right)
 \prod_{k=1}^{\infty} \left| 
 1+e^{ 2\pi i k  \left(  
   \tau_1 
   +i \tau_2 \sqrt{ 1+ \left(\frac{\mu}{2\pi k} \right)^2}
   \right) }
 \right|^2, \\
Z_4^{\rm cont.} &= 
 \left(
 \prod_{k \in {\mathbb Z}} 
 e^{ \pi \tau_2 \sqrt{ k^2 + \left(\frac{\mu}{2\pi}\right)^2 }  }
 \right)
 \left(
 1-e^{-\tau_2 \mu} 
 \right)
 \prod_{k=1}^{\infty} \left| 
 1-e^{ 2\pi i k  \left(  
   \tau_1 
   +i \tau_2 \sqrt{ 1+ \left(\frac{\mu}{2\pi k} \right)^2}
   \right) }
 \right|^2. \\
\end{split}
\end{equation}
which exactly agree\footnote{
The prefactor in E.(\ref{Ising PF}) gives 
a non-universal overall factor in the continuum limit. 
} with the result in Eq.(\ref{MF Z}) of continuum 
field theory of a free massive Majorana fermion 
\begin{equation}
Z_1^{\rm cont.} = Z_{\frac{1}{2},\frac{1}{2}}, \; 
Z_2^{\rm cont.} = Z_{0,\frac{1}{2}}, \; 
Z_3^{\rm cont.} = Z_{\frac{1}{2},0}, \; 
Z_4^{\rm cont.} = Z_{0,0}, 
\end{equation}
where the labels $\mu,\nu=0, \frac{1}{2}$ of $Z_{\mu,\nu}$ 
specify the boundary condition of the fermion as described 
in Appendix \ref{app:massive fermion}. 
This means that the continuum limit of the 
two-dimensional Ising model with the boundary condition in 
Eq.(\ref{bdr cond}) is the two-dimensional massive fermion theory 
on the torus. 
In particular, the combination $\frac{p}{n}+i\frac{m}{n}$ becomes
the complex structure $\tau=\tau_1+i\tau_2$ of the continuum torus 
as shown in Eq.(\ref{two moduli}), and the parameter $\mu$ 
in Eq.(\ref{eq:deviation_critical_temp}) representing the 
deviation from the critical temperature is nothing but the 
mass parameter of the fermion in the continuum limit. 

\begin{figure}[htbp]
\begin{center}
\scalebox{0.4}{
\includegraphics{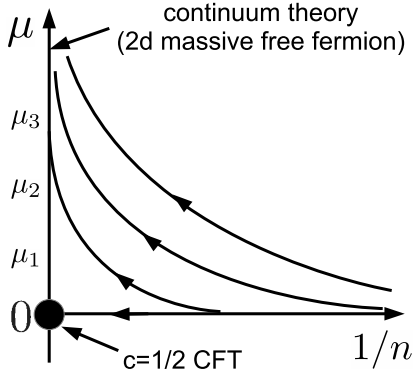}}
\caption{The parameter space of the 2D Ising model with the condition 
$p=\tau_1 n$ and $m=\tau_2 n$. 
The curved lines express the RG flows of an irrelevant operator 
associated with $\frac{1}{n}$. 
After taking the continuum limit, 
the parameter $\mu$ is identical to the mass of the two-dimensional 
free fermion. }
\label{fig:RG flow}
\end{center}
\end{figure}

Before closing this section, it is worth looking at the RG 
structure of the Ising model. 
Fixing parameters $\tau_1$ and $\tau_2$ in Eq.(\ref{two moduli}) 
which specify the geometry of the torus, the parameter space 
of the Ising model is spanned by $(n,\mu)$, 
and the procedure of taking the continuum limit is nothing but 
taking the limit of $n\to\infty$ with fixing $\mu$. 
As we have seen, 
after taking this limit, the parameter $\mu$ is precisely identical 
to the mass parameter of the two-dimensional massive free fermion
and the limit $\mu\to 0$ corresponds to the massless 
Majorana fermion, that is, $c=\frac{1}{2}$ CFT. 
Therefore the flow parametrized by the parameter $\mu$ starting from the 
conformal fixed point exactly equals to the mass deformation
of the $c=\frac{1}{2}$ CFT (Fig.~\ref{fig:RG flow}). 


\section{Holographic description of the RG flow}
\label{sc:holgraphicRGflow}

\subsection{Holographic RG flow of single scalar field
}
\label{sc:RG_scalar}

In this section, we holographically describe 
the RG flow starting from 
the conformal fixed point along the parameter $\mu$.

In the spirit of the AdS/CFT correspondence, the source 
(coupling constant) of an operator in the boundary field theory 
is identified with a field in the bulk gravity. 
The value of the coupling constant of the boundary field theory varies 
towards infrared (IR) as a RG flow. 
On the other hand, the classical solution of the bulk gravity 
provides a trajectory of the bulk field along 
the radial direction of an asymptotic AdS geometry. 
This trajectory is regarded as the RG flow of the coupling 
constant of the boundary field theory away from a CFT at 
ultraviolet (UV) fixed point. 
As mentioned in the introduction, 
the two-dimensional $c=\frac{1}{2}$ CFT is conjectured to be 
dual to the ``quantum'' pure gravity under the assumption that 
the path integral over the metric of 3D space-time is 
localized to the classical solutions, that is, the BTZ black holes. 
We here assume that the same is true 
at least in the neighborhood of the CFT fixed point; 
the classical solutions are dominant even in the presence 
of additional fields in evaluating the partition function of 
the quantum gravity. 

In the analysis of the previous section, we found that 
there are two independent flows from the conformal fixed point 
parametrized by $\mu$ and $1/n$. 
We expect that the parameter $1/n$ in the boundary theory 
corresponds to certain discretized version of bulk gravity, 
which is difficult to work out at present. 
In the following, we explore the continuum bulk gravity to 
study the RG flow of the parameter $\mu$, which can be 
identified as the mass of the fermion of the boundary field 
theory. 

Since the parameter $\mu$ couples to the operator $\bar\Psi \Psi$ 
at the boundary field theory, it is plausible to assume that 
the corresponding field in the gravity is a real scalar field 
$\phi$, which may be considered as the minimum number of 
degrees of freedom to describe the RG flow of a single 
parameter $\mu$. 
We thus consider the following action of three dimensional 
Euclidean gravity with a real scalar field $\phi$; 
\footnote{We here simply omit to write the boundary terms of 
the gravity action.} 
\begin{equation}
S=\frac{1}{16\pi G_{\rm N}}\int d^3 x \sqrt{g}
\left[-R+V(\phi)+\frac{1}{2}g^{\mu\nu}K(\phi) \partial_\mu \phi 
\partial_\nu \phi\right],
\label{eq:scalar_gravity_action0}
\end{equation}
where 
$G_N$ is the Newton constant of three-dimensional gravity, 
$g_{\mu\nu}$ $(\mu,\nu=1,2,3)$ is the metric, 
$R$ is the Ricci scalar, 
$g={\rm det} (g_{\mu\nu})$, 
and $K(\phi)$ is a function of $\phi$, which describes the 
nonlinearity of the kinetic term of scalar field $\phi$. 
Since the AdS$_3$ geometry with the radius $L$ should be a 
solution of this system when $\phi=0$, we demand $V(\phi)$ 
to satisfy 
\begin{equation}
V(\phi=0)=-\frac{2}{L^2}. 
\label{eq:potential_origin}
\end{equation}
Since $\phi$ should have a nonsingular kinetic term at least 
for small $\phi$, we further require that $K(\phi)$ is 
regular at $\phi=0$. Without loss of generality, we can 
fix 
\begin{equation}
K(\phi=0)=1 
\label{eq:normalization_K}
\end{equation}
by choosing the normalization of the field $\phi$. 
Our final task is to determine functions $V(\phi)$ and $K(\phi)$ 
by requiring the solutions $\phi, g_{\mu\nu}$ to describe the 
holographic RG flow of the Ising model off critical 
temperature.

\subsection{Solution of the gravity corresponding to the mass deformation}
\label{sc:ads_solutions}

Since we are interested in the evolution of the scalar field 
along the radial direction of an asymptotic AdS geometry, 
we set the following ansatz for the metric and the scalar field:
\begin{equation}
ds^2=e^{2h(r)} dr^2 + \frac{r^2}{L^2} \sum_{i=1,2} d x_i^2, \qquad
\phi=\phi(r), 
\label{ansatz1}
\end{equation}
where $x_i$ $(i=1,2)$ express the two-dimensional transverse 
directions, and $r\in[0,\infty)$ is the radial coordinate, 
which may be regarded as the Euclidean time. 
We have also assumed that the functions $h(r)$ and $\phi(r)$ 
depend only on $r$. 
Note that we have fixed the gauge by setting 
$g_{ij}=\frac{r^2}{L^2}\delta_{ij}$ in Eq.(\ref{ansatz1}). 
Since the geometry is asymptotically anti de Sitter space, 
$e^{2h}$ must be expandable around $r=\infty$ as 
\begin{equation}
e^{2h}=\frac{L^2}{r^2}+\cO(r^{-4}). 
\label{asympt}
\end{equation}
In this setup, the independent field equations are given by%
\footnote{
The field equation of the scalar matter $\phi$ can be derived from 
those of the metric.
}
\begin{align}
\label{eq1}
&\frac{1}{2} K(\phi) \dot{\phi}(r)^2 = \frac{1}{r^2} 
+ \frac{\dot{h}(r)}{r}, \\
&V(\phi)=e^{-2h(r)} \left( -\frac{1}{r^2} +  \frac{\dot{h}(r)}{r} \right),
\label{eq2}
\end{align}
where the dot $\dot{}$ expresses the derivation with respect 
to $r$. 
Our first task is to obtain $h(r)$ and $\phi(r)$ for given fixed 
functions $K(\phi)$ and $V(\phi)$ by solving Eqs. (\ref{eq1}) 
and (\ref{eq2}). 
{}Using the boundary condition in Eq.(\ref{asympt}), we can 
integrate Eq.(\ref{eq1}) to obtain $h(r)$ in terms of 
$\phi(r)$ and $K(\phi)$ as 
\begin{equation}
e^{2h(r)}= \frac{L^2}{r^2}e^{-\int_r^{\infty} sK(\phi(s))(\dot\phi(s))^2 ds}. 
\label{solution h}
\end{equation}
The standard way to solve the equations would be 
plugging this to Eq.(\ref{eq2}) and solve the obtained single equation 
of $\phi(r)$ for given $K(\phi)$ and $V(\phi)$. 
However we proceed the opposite way: 
We first fix the behavior of $\phi(r)$ from a physical requirement 
and determine the relation between $K(\phi)$ and $V(\phi)$.

Since the mass deformation keeps the theory free, 
the scaling of the mass parameter should be trivial. 
Then we can assume 
\begin{equation}
\mu(a) = \frac{a_0}{a}\mu_0, 
\label{mass scale}
\end{equation}
where $a$ is the length scale of the (boundary) field theory 
and $\mu_0$ is the mass at the reference length scale $a_0$. 
We should also recall that, in the context of the holographic 
RG, the radial coordinate $r$ is identified with the RG 
parameter, namely the ``Euclidean time''. 
The evolution of a bulk field along this radial direction 
is identified with the RG flow of the corresponding 
coupling constant. 
In our choice of gauge in Eq.(\ref{ansatz1}), 
the radial coordinate $r$ can be identified as the length 
scale $a$ of the two-dimensional boundary field theory. 
Then the behavior of the field $\phi(r)$ can be regarded as 
the scaling behavior of the corresponding coupling constant 
\cite{de Boer:1999xf,Fukuma:2000bz}.  
In the present context, we are looking for such a solution 
that corresponds to the mass parameter $\mu$ of the free fermion 
theory which exactly behaves as Eq.(\ref{mass scale}). 
To achieve this goal, 
we should fix the solution of the scalar field as 
a function of radial coordinate $r$ (length scale parameter) 
\begin{equation}
\phi(r) = \frac{L}{r} \phi_0 \equiv \bar{\phi}(r)
\label{ansatz2}
\end{equation}
with a constant $\phi_0$, instead of solving $\phi(r)$ 
for given fixed $K(\phi)$ and $V(\phi)$. 
Eq.(\ref{ansatz2}) allows us to change a variable from $r$ 
to $\phi$, and Eq.(\ref{solution h}) becomes 
\begin{equation}
e^{2h(r)}=\frac{L^2}{r^2}e^{-\int_0^{\bar\phi(r)} xK(x) dx}. 
\label{solution h2}
\end{equation}
Substituting Eqs.(\ref{ansatz2}) and (\ref{solution h2}) into 
Eq.(\ref{eq2}), we obtain a relation between $K(\phi)$ and 
$V(\phi)$  
\begin{equation}
V(\phi)=-\frac{2}{L^2}\left( 1 - \frac{\phi^2}{4} K(\phi) \right)
e^{\int_0^{\phi} xK(x) dx}.
\label{KandV}
\end{equation}
This is a necessary condition for the bulk scalar field $\phi$ 
to become the holographic dual to the mass parameter of the 
free fermion.

\subsection{Holographic RG flow in terms of Hamilton-Jacobi equation}
\label{sc:hamilton_Jacobi}

Even after imposing the exact scaling behavior of the mass 
parameter in Eqs.(\ref{mass scale}) and (\ref{ansatz2}), we 
still have a freedom to choose one function of 
$\phi$ in the action, either $K(\phi)$ or $V(\phi)$, in order 
to represent the holographic RG flow of our system. 
In fact, with the requirement of the scalar field behavior 
$\phi(r)$ in Eq.(\ref{ansatz2}), the bulk geometry is uniquely
determined by Eq.(\ref{solution h}) for any choice of either 
$K(\phi)$ or $V(\phi)$ which are related by Eq.(\ref{KandV}) 
as a consequence of field equations.

In order to fix it, let us further consider the holographic 
RG structure of this system based on the Hamilton-Jacobi equation 
of the bulk gravity%
\cite{de Boer:1999xf,Fukuma:2000bz}: 
\begin{align}
G_{ij:kl} 
\left( \frac{1}{\sqrt{g}} \frac{\delta S_{\rm cl}}{\delta g_{ij}} \right)
\left(\frac{1} {\sqrt{g}} \frac{\delta S_{\rm cl}}{\delta g_{kl}} \right)
+\frac{1}{2K(\phi)} \left(\frac{1} {\sqrt{g}} 
\frac{\delta S_{\rm cl}}{\delta \phi} \right)^2 
= V(\phi)-R+\frac{K(\phi)}{2}g^{ij}\del_i \phi \del_j \phi,
\label{HJ eq}
\end{align}
where $G_{ij;kl}$ is defined by 
$G_{ij;kl}\equiv g_{ik}g_{jl}-g_{ij}g_{kl}$, and $R$ is 
the two-dimensional scalar curvature constructed from $g_{ij}(x)$. 
The classical action $S_{\rm cl}=S_{\rm cl}[g_{ij}(x),\phi(x)]$ 
is a functional of the boundary values of the metric 
$g_{ij}(x)$ $(i,j=1,2)$ and the scalar field $\phi(x)$ on the 
two-dimensional surface at a specific value of the radial 
coordinate, say $r=r_0$, 
and is obtained by substituting the classical solution 
into the bulk action in Eq.(\ref{eq:scalar_gravity_action0}).

The momentum constraint of the bulk gravity insures that the 
classical action is invariant under the diffeomorphism of 
the two-dimensional boundary. 
We can then expand the classical action $S_{\rm cl}$ 
in powers of derivatives%
\footnote{
In Refs.\cite{de Boer:1999xf,Fukuma:2000bz}, the authors divide the 
classical action into the local and the non-local parts as 
$S[\phi(x),g_{ij}(x)]=S_{\rm loc} [\phi(x),g_{ij}(x)] 
+ \Gamma[\phi(x),g_{ij}(x)]$, 
and show that the non-local part 
$\Gamma$ satisfies the RG equation of the boundary field theory. 
In this paper, we further expand $\Gamma[\phi(x),g_{ij}(x)]$ 
in powers of derivatives. 
} 
in the transverse directions: 
\begin{align}
S_{\rm cl} = \int d^2x \sqrt{g} \left( W(\phi) + \cdots
\right), 
\label{classical S}
\end{align}
where $\cdots$ includes terms with the derivatives of $\phi$ 
and the two-dimensional curvature tensors, which vanish when 
$\phi$ is independent of $x^i$ and the transverse geometry is 
flat. 
The Hamilton-Jacobi equation in Eq.(\ref{HJ eq}) can also be 
expanded in powers of derivatives, 
and gives the following relation among $W(\phi)$, $K(\phi)$ 
and $V(\phi)$ at the leading order of the expansion in powers 
of derivatives\cite{de Boer:1999xf,Fukuma:2000bz} 
\begin{equation}
V(\phi)=
-\frac{1}{2} W(\phi)^2 +\frac{1}{2K(\phi)}
W'(\phi)^2. 
\label{eq:W_V_relation}
\end{equation}
By combining this relation with Eq.(\ref{KandV}), we can 
determine $W(\phi)$ in terms of $K(\phi)$ as 
\begin{equation}
W(\phi)=\frac{2}{L}\exp\left({\frac{1}{2}\int_0^\phi 
xK(x) dx}\right). 
\label{eq:W_K_relation}
\end{equation}
Also, 
repeating  the argument given in Refs.~\cite{de Boer:1999xf,Fukuma:2000bz}, 
we can obtain the $\beta$-function of $\phi$ as
\begin{equation}
\beta(\phi)=
\frac{2}{K(\phi)} \frac{ W'(\phi) }{ W(\phi) }, 
\label{eq:beta-function}
\end{equation}
and the holographic c-function 
\begin{equation}
c(\phi)= \frac{3}{G_N} \frac{1}{ W(\phi)}, 
\label{eq:c-function}
\end{equation}
which is a monotonically decreasing function when the coefficient of the 
kinetic term $K(\phi)$ is positive definite. 
Note that the c-function (\ref{eq:c-function}) is defined as the coefficient 
of the scalar curvature appearing in evaluating the expectation value 
of the energy-momentum tensor.  
This kind of the c-function has also been proposed 
using functional renormalization group equation\cite{Codello:2013iqa}.

The gauge/gravity correspondence asserts that the classical action of 
the bulk gravity is regarded as the (regularized) free energy of the dual
boundary field theory. 
In our case, we have considered the three-dimensional gravity 
with a single bulk scalar field $\phi$ as a candidate of 
the dual description of the massive two-dimensional free fermion. 
Therefore one may naively expect that $e^{-S_{\rm cl}}$ should be equal 
to the partition function of the two-dimensional massive free 
fermion in Eq.(\ref{cont PF}), 
when $\phi$ is independent of the transverse coordinates $x^i$ 
and the transverse geometry is flat.
However, 
we should recall that the Ising field theory corresponds 
to the three-dimensional ``quantum'' gravity 
\cite{Witten:2007kt,Maloney:2007ud}. 
The three-dimensional gravity with appropriate boundary 
conditions possesses the conformal symmetry with the central 
charge $\frac{1}{2}$ at the boundary\cite{Brown:1986nw}. 
In Ref.~\cite{Castro:2011zq}, as mentioned in Introduction, 
the three-dimensional pure gravity is considered. 
It has been show that the integration over all the possible three-dimensional 
metric with fixing boundary condition is localized to the classical
three-dimensional geometries  (BTZ black holes) 
and the partition function of the quantum gravity turns out to 
that of the Ising field theory by relying on the boundary conformal symmetry. 
Our analysis in this paper is based on the same assumption that 
this quasi-semi-classical approach still makes sense 
even after adding the mass term to the boundary field theory. 
Therefore, 
the partition function of the massive fermion as a sum of 
functions of the moduli parameter $\tau$ in Eq.(\ref{cont PF})
should be obtained after summing up all the possible 
classical geometries.

However it is actually hard to carry out this procedure 
explicitly, since we look at only one solution of the 
classical field equations and the conformal symmetry is broken
by introducing nonvanishing values of the scalar field $\phi$ 
to represent the mass term of the boundary fermion. 
However it is still possible to determine the scalar potential 
$V(\phi)$ of the bulk gravity as follows. 
We observe that the boundary condition of the fermion becomes 
irrelevant and that the partition function becomes $\tau$-independent 
when the geometry of the boundary becomes $\R^2$. 
In this case, the partition function becomes extremely simple and 
we can expect that every solution gives the same contribution 
to the partition function even if there are several solutions 
in the bulk with the same boundary condition. 
When the geometry is $\R^2$, the free energy of the massive 
fermion is given by 
\begin{equation}
{\mathcal F}= \int \frac{ d^2p }{(2\pi)^2}  
\log \left( p^2 + \mu^2 \right), 
\label{eq:free_energy_free_massive_fermion}
\end{equation}
where $\mu$ is the mass parameter. 
In order for the free energy in 
Eq.(\ref{eq:free_energy_free_massive_fermion}) to be well-defined, 
we need to regularize both UV and IR divergences. 
Although the result depends on details of the regularization, 
the regularized free energy in general takes the form 
\begin{equation}
{\mathcal F}_{\rm reg} = a + b \mu^2 + c\mu^2 \log \mu, 
\label{reg free energy}
\end{equation}
where $a$, $b$ and $c$ are some constants.

As mentioned above, we regard the scalar field $\phi$ at the 
boundary as the mass parameter of the boundary fermion 
and we identify the regularized free energy with the 
classical action $S_{\rm cl}$ in Eq.(\ref{classical S}) 
at the boundary which is obtained from the bulk three-dimensional 
gravity. 
In addition, $W(\phi)$ in Eq.(\ref{classical S}) is 
proportional to the free energy 
because $\phi$ does not have $x^i$-dependence. 
Therefore $W(\phi)$ can be expressed as 
\begin{equation}
W(\phi) = A + B \phi^2 + C\phi^2 \log \phi, 
\label{free energy}
\end{equation}
where $A$, $B$ and $C$ are some constants. 
Using Eqs.(\ref{eq:W_K_relation}) and  (\ref{eq:W_V_relation}), 
we can determine $K(\phi)$ and $V(\phi)$ in the following. 

\begin{figure}[htbp]
\begin{center}
\scalebox{0.5}{
\includegraphics{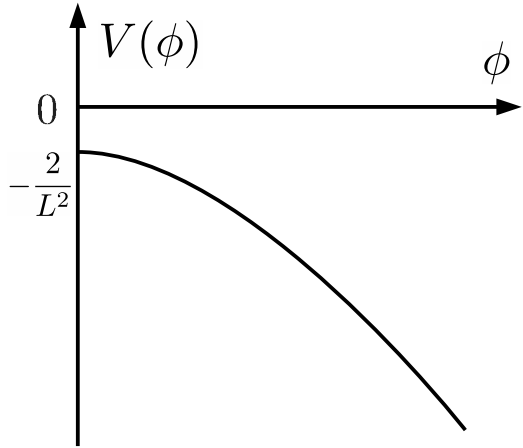}}
\caption{
The shape of the potential 
corresponding to the flow from the $c=\frac{1}{2}$ CFT to 
the infrared by the mass deformation. 
The point $\phi=0$ corresponds to the AdS boundary with the radius $L$ 
and $\phi=\infty$ corresponds to $r=0$. 
}
\label{fig:solution}
\end{center}
\end{figure}

Suppose $C\ne 0$, we observe that $W(\phi)$ is dominated by 
the last term $C\phi^2 \log \phi$ at small values of $\phi$. 
This implies that the RG flow at $\phi \to 0$ is discontinuous 
with the AdS solution in Eq.(\ref{eq:W_K_relation}) when $C\ne 0$. 
Since we identify the evolution of $\phi$ from $r=0$ to $r=\infty$ 
along the radial direction (the Euclidean "time evolution") 
as the RG flow from the UV fixed point to IR, 
this behavior is not acceptable as the RG flow of the mass parameter. 
Therefore, we should adopt a regularization scheme in the 
holographic renormalization group by imposing the condition 
\begin{equation}
c=C=0. 
\end{equation}
Note that the dimensional regularization realizes this condition 
for example. 
With this choice of $W(\phi)$, we obtain $K(\phi)$ and $W(\phi)$ 
from Eq.(\ref{eq:W_K_relation}) and the normalization condition 
Eq.(\ref{eq:normalization_K}), 
\begin{equation}
K(\phi)=\frac{1}{1+\frac{1}{4}\phi^2}, 
\label{eq:K}
\end{equation}
and 
\begin{equation}
W(\phi)=\frac{2}{L}\left(1+\frac{1}{4}\phi^2\right).
\label{eq:W}
\end{equation}
Incidentally, we obtain the $\beta$-function (\ref{eq:beta-function}) as 
\begin{equation}
\beta(\phi)=\phi, 
\label{eq:our beta}
\end{equation}
and the c-function from (\ref{eq:c-function}) as 
\begin{equation}
c(\phi)=\frac{1}{2+\frac{\phi^2}{2}}. 
\label{eq:our c}
\end{equation}
The beta-function (\ref{eq:our beta}) is consistent with the 
notion that the bulk scalar field $\phi$ 
corresponds to the mass parameter of the boundary field theory, 
and the c-function (\ref{eq:our c}) 
is a monotonically decreasing function of $\phi\in[0,\infty)$ 
from $\frac{1}{2}$ to $0$ as expected. 
Substituting this result into Eq.(\ref{KandV}), we obtain 
\begin{equation}
V(\phi)=-\frac{2}{L^2}\left( 1 + \frac{\phi^2}{4} \right), 
\label{eq:potential}
\end{equation}
which is monotonically decreasing $V\to -\infty$ as 
$\phi\to\infty$ (see fig.\ref{fig:solution}). 
This is also consistent with the expectation that mass perturbation 
deforms the $c=1/2$ Ising CFT to flow to a system with less 
degrees of freedom as dictated by the c-theorem. 
Namely it is likely to flow to nothing ($c=0$) in the IR. 

\section{Summary and Discussion}
\label{sc:conclusion_discussion}

In this article, we have obtained the partition function of 
the Ising model on Euclidean two-dimensional lattice with 
the twisted boundary condition representing the torus with 
a complex structure $\tau$ in a discretized version. 
We have taken an appropriate scaling limit to obtain a continuum 
limit for the torus with the complex structure $\tau$, retaining 
a deviation from the critical temperature. 
The resulting continuum partition function agrees with that 
of the mass-deformed Ising CFT, namely the continuum field 
theory of massive Majorana fermion on the torus with the 
complex structure $\tau$. 
We have also discussed the RG flow of the Ising model off 
critical temperature in terms of the three-dimensional AdS 
gravity together with a scalar field $\phi$ representing 
the RG flow of the mass parameter. 
By using the Hamilton-Jacobi equation, we have found that 
a simple nonlinear kinetic function $K(\phi)$ in 
Eq.(\ref{eq:K}) and a simple potential $V(\phi)$ in 
Eq.(\ref{eq:potential}) for the scalar field $\phi$ 
describes the expected RG flow from $c=1/2$ CFT at the UV 
region towards $c=0$ at the IR.

Let us comment on the relation of our results to those in 
Ref.\cite{Hotta:2008xt} where a similar setup has been used 
to look for solutions interpolating two AdS geometries 
corresponding to two different conformal field theories. 
They assumed the canonical kinetic term $K(\phi)=1$ 
for the scalar field $\phi$, which is achieved 
by a field redefinition from our case.
They required that the metric component $G_{tt}$ has a 
double zero at the horizon $r_0$ (IR), at which the metric 
becomes another AdS geometry in addition to an AdS geometry 
at $r=\infty$ (UV). 
Near $r \to \infty$ at UV, their solution exhibits a chirally 
asymmetric Virasoro algebra, where the excitation spectra 
of the left and right Virasoro algebras, $L_0$ and $\bar L_0$ 
are different. 
This peculiar behavior comes from their Ansatz with the 
non-vanishing metric component $G_{t\varphi}$, where $t, \varphi$ 
are coordinates other than $r$ corresponding to our $x_1, x_2$ 
in Eq.(\ref{ansatz1}). 
We have chosen a different ansatz with $G_{t\varphi}=0$ 
in Eq.(\ref{ansatz1}) to obtain chirally symmetric Virasoro 
algebras, as in the Ising model.

We have used the Hamilton-Jacobi equation to describe the RG 
flow of partition functions of the Ising model on the torus. 
There are more detailed informations such as multi-point 
correlation functions or (equivalently) partition functions 
on higher genus Riemann surfaces. 
In principle, one should be able to describe these data, 
which are worth studying. 
However, it is likely that one needs to overcome the problems 
associated with the strong coupling, or quantum effects in 
gravity in order to describe these data quantitatively 
in the presence of matter fields.

Recent applications of AdS/CFT to condensed matter physics 
provide many interesting insights into possible phase 
structures of strongly coupled system. 
In particular, introducing periodicity into the system to 
mimic lattice structures seems to be a crucial feature of 
these applications\cite{nernst,Hartnoll:2012rj,Blake:2013owa,
Blake:2014yla,Baggioli:2014roa}. 
Since many realistic models of statistical physics or condensed 
matter physics are built on some discretized system, it is 
desirable to obtain discretized models as the boundary of some 
discretized model for the bulk in discussing the AdS/CFT 
correspondence. 
We hope that the continuum space-time would emerge by taking 
the continuum limit of the discretized system in the bulk 
and the gravity would be realized as a sort of a cooperative 
phenomenon. 
This ambitious objective is of course still quite hard to achieve. 
How to obtain a discretized version of (quantum) gravity 
realizing the given discretized model on the boundary 
is the key question which is worth pursuing. 
With this philosophy in mind and taking our result quite 
optimistic, we may expect that the usual two-dimensional 
Ising model would be a candidate of the discrete boundary 
model related to a discrete version of quantum gravity in the 
above sense, and our result may be a first step toward this 
direction, since the deviation parameter $\mu$ from the 
critical temperature is actually included in the original 
discretized Ising model. 
However our discussion still remained within the continuum theory. 
In this sense, it would be a great step if one finds a gravity 
description of the RG flow parametrized by $\frac{1}{n}$ in 
Fig.~\ref{fig:RG flow}. 
Such an attempt is interesting in its own right, and furthermore 
can provide a concrete starting point to incorporate 
lattice structures into the AdS/CFT correspondence, which may 
play a vital role in condensed matter physics applications.

\section*{Acknowledgment}

We thank Tetsuo Deguchi for a valuable discussion on the Ising 
model, 
and Yoshifumi Hyakutake and Keiju Murata 
for useful discussions on the RG flow 
in the context of the AdS/CFT correspondence. 
This work is supported in part by Grant-in-Aid for 
Scientific Research from the Ministry of Education, Culture, 
Sports, Science and Technology, Japan No.25400271 (N.S.), 
and Grant-in-Aid for Scientific Research (C), 15K05060 (S.M.). 

\appendix
\section{Transformation of $H_\pm$ and $\Sigma_\pm$}
\label{app:trans}

Let us express the generators of $Spin(2n)$ in the fundamental representation as 
\begin{equation}
\left(\hat{J}_{\mu\nu}\right)_{\rho\sigma} = i \left( \delta_{\mu\rho}\delta_{\nu\sigma}
-\delta_{\mu\sigma}\delta_{\nu\rho} \right).\qquad 
(1\le \mu < \nu \le 2n)
\end{equation}
Let us consider $H_\pm$ and $\Sigma_\pm$ defined in (\ref{def Hpm}) 
in the fundamental representation, 
which are matrices with the size of $2n$ and we write them 
as $\hat{H}_\pm$ and $\hat{\Sigma}_\pm$, respectively, in the following. 
They are written as 
\begin{equation}
\begin{split}
\hat{H}_\pm=\left( \begin{matrix} 
x               & y &       0 & \cdots  &   0 &  \mp y^\dagger \\
y^\dagger & x & y & 0 &  \cdots & 0 \\
0 & y^\dagger & x & y & \cdots & 0 \\
\vdots &    &               &\ddots  &  & \vdots \\
\vdots &    &      &         &\ddots   & \vdots \\
\mp y    & 0 & \cdots          &    & y^\dagger & x
\end{matrix}\right) , \quad 
\hat\Sigma_\pm=
\alpha_\pm
\left( \begin{matrix} 
0 & 0 & 0 &  \cdots & \mp{\mathbf 1}_2 \\
\mathbf{ 1}_2 & 0 & 0 &  \cdots & 0 \\
0 & \mathbf {1}_2 & 0 &  \cdots & 0 \\
\vdots &   & \ddots & &  \vdots \\ 
0 & 0 & \cdots & \mathbf {1}_2 &  0 
\end{matrix}\right), 
\end{split}
\end{equation} 
where $x$ and $y$ are $2\times 2$ matrices, 
\begin{equation}
x\equiv \left(\begin{matrix}
\cosh 2\ta \cosh 2b & i \sinh 2\ta \cosh 2b \\
-i \sinh 2\ta \cosh 2b & \cosh 2\ta \cosh 2b 
\end{matrix}\right), \quad 
y\equiv \left(\begin{matrix}
-\frac{1}{2} \sinh 2\ta \sinh 2b & -i \sinh^2 \ta \sinh 2b \\
i \cosh^2 \ta \sinh 2b & -\frac{1}{2} \sinh 2\ta \sinh 2b 
\end{matrix}\right), 
\end{equation}
and the overall factor $\alpha_\pm$ in the definition of $\hat\Sigma_{\pm}$ 
is given by 
\begin{equation}
\alpha_+=
\begin{cases}
1 & (n:{\rm even}) \\
-i & (n:{\rm odd})
\end{cases}, \quad
\alpha_- =
\begin{cases}
-i & (n:{\rm even}) \\
1 & (n:{\rm odd})
\end{cases}, 
\end{equation}
which are necessary in order to reproduce (\ref{shift matrix in spin}) 
in the spin representation. 

Our goal in this appendix is explicitly transforming the matrices
$\hat{H}_{\pm}$ and $\hat\Sigma_\pm$ 
into this canonical form. 
Then we can easily estimate ${\rm Tr}_\pm H_\pm^m \Sigma_\pm^p$ 
in the spin representation by using (\ref{trace}).

We first introduce matrices $\Omega_\pm \in SO(2n)$ whose $(i,j)$ blocks are given by 
\begin{equation}
\left(\Omega_+\right)_{ij}= \frac{1}{\sqrt{n}} R\left(\frac{2\pi\, i(j-1/2)}{n} \right), \quad 
\left(\Omega_-\right)_{ij}=\frac{1}{\sqrt{n}} R\left(\frac{2\pi\, ij}{n}\right), 
\qquad (i,j=1,\cdots,n)
\end{equation}
respectively, where $R(\theta)$ is defined in (\ref{2d rotation}). 
The similarity transformations of $H_\pm$ and $\Sigma_\pm$ 
by $\Omega_\pm$ become
\begin{align}
\begin{split}
\Omega_+^T \hat{H}_+ \Omega_+ 
&=\left(\begin{matrix}
M_{1} & & & & & N_{2n-1} \\
 & M_{3} & & & N_{2n-3} \\
 & & \ddots  & \iddots     \\         
 & & \iddots  & \ddots     \\         
 & N_{3} & & & M_{2n-3} \\
N_{1} & & & & & M_{2n-1} \\
\end{matrix}\right), \\
\Omega_+^{T} \hat\Sigma_+ \Omega_+ 
&=\left(\begin{matrix}
R\left(-\frac{\pi}{n}\right) & \\
& R\left(-\frac{3\pi}{n}\right) & \\
& & \ddots &  \\
& & & R\left(\frac{3\pi}{n}\right)  \\
& & & &  R\left(\frac{\pi}{n}\right) & 
\end{matrix}\right), \\
\end{split}
\label{matrix+1}
\end{align}
and
\begin{align}
\begin{split}
\Omega_-^{T} \hat{H}_- \Omega_- 
&=\left(\begin{matrix}
M_{2} & & & & & N_{2n-2} & 0 \\
 & M_{4} & & & N_{2n-4} \\
 & & \ddots  & \iddots     \\         
 & & \iddots  & \ddots     \\         
 & N_{4} & & & M_{2n-4} \\
N_{2} & & & & & M_{2n-2} & 0 \\
 0 & & & & & 0 & M_{0} 
\end{matrix}\right), \\ 
\Omega_-^{T} \hat\Sigma_- \Omega_- 
&=\left(\begin{matrix}
R\left(-\frac{2\pi}{n} \right) & \\
& R\left(-\frac{4\pi}{n}\right) & \\
& & \ddots  \\
& & & R\left(\frac{4\pi}{n}\right) & \\
& & & & R\left(\frac{2\pi}{n}\right) \\
& & & & & {\mathbf 1}_2
\end{matrix}\right), 
\end{split}
\label{matrix-1}
\end{align}
where $M_{I}$ and $N_{{I}}$ $(I=1,\cdots,2n)$ are matrices, 
\begin{equation}
M_{I} \equiv
\left(\begin{matrix}
 A_I & iB_I \\
 -iB_I & A_I 
\end{matrix}\right), \quad
N_{I}\equiv 
\left(\begin{matrix}
i C_I & 0 \\
0 & -i C_I
\end{matrix}\right), 
\label{tochu}
\end{equation}
with 
\begin{align}
\begin{split}
A_I &= \cosh 2\tilde{a} \cosh 2b - \cos\left(\frac{\pi I}{n}\right) \sinh 2\tilde{a} \sinh 2b, \\
B_I &= \sinh 2\tilde{a} \cosh 2b - \cos\left(\frac{\pi I}{n}\right) \cosh 2\tilde{a} \sinh 2b, \\
C_I &= \sin\left(\frac{\pi I}{n}\right) \sinh 2b. 
\end{split}
\end{align}
Note that $I$ runs odd (even) numbers for the matrices with the index $+$ ($-$).
It is easy to see that $M_I$ and $N_I$ satisfy
\begin{equation}
M_{2n-I}=M_I, \quad N_{2n-I}=-N_I. 
\end{equation}

Since $A_I$, $B_I$ and $C_I$ satisfy 
\begin{equation}
A_I^2 - B_I^2 - C_I^2 = 1, 
\end{equation}
we can uniquely determine the parameters $\gamma_I>0$, $\theta_I\in[0,\frac{\pi}{2}]$ 
and $\epsilon_r=\pm 1$ by 
\begin{equation}
A_I \equiv \cosh\gamma_I, \quad
B_I \equiv \epsilon_I \sinh\gamma_I \cos\theta_I, \quad 
C_I \equiv \pm \sinh\gamma_I \sin\theta_I, 
\label{ABC}
\end{equation}
where the sign in the definition of 
$C_I$ takes $+$ for $1\le I \le n$ and $-$ for $n+1 \le I \le 2n$. 
Note that the $\gamma_I$ appearing in (\ref{ABC}) is the same one defined in (\ref{gamma-r}). 
We also note that $N_0=N_n=0$ and $M_0$ and $M_n$ are given by 
\begin{equation}
M_0=R(\pm i \gamma_0), \quad
M_n=R(-i\gamma_n), 
\label{M0 and Mn}
\end{equation}
where the sign appearing in the expression of 
$M_0$ takes $+$ in the disordered phase $(\tilde{a}>b)$ 
and $-$ in the ordered phase $(\tilde{a}<b)$.

We can rearrange the matrices (\ref{matrix+1}) and (\ref{matrix-1}) by permuting the elements 
properly. To this end, we introduce the matrices which generate the transportations, 
\begin{equation}
T_{i,j}\equiv\left(\begin{matrix}
{\mathbf 1}_{2(i-1)} & \\
& 0_2 &\hdots& {\mathbf 1}_2  \\
&\vdots& {\mathbf 1}_{2(j-i-1)} & \vdots \\ 
& {\mathbf 1}_2 & \hdots & 0_2 \\
&&&& {\mathbf 1}_{2(n-j)} \\
\end{matrix}\right), \quad (1\le i < j \le n)
\end{equation}
and the cyclic rotations, 
\begin{equation}
C_{i,j} \equiv T_{j-1,j}T_{j-2,j-1}\cdots T_{i,i+1}. 
\end{equation}
Then by defining 
\begin{align}
\begin{split}
S_+ &= \begin{cases}
C_{2,n} C_{4,n} \cdots C_{n-2,n} & (n:{\rm even}) \\
C_{\frac{n+1}{2},n}^{-1} C_{2,n-1} C_{4,n-1} \cdots C_{n-3,n-1} & (n:{\rm odd}) 
\end{cases}, \\
S_- &= \begin{cases}
C_{\frac{n}{2},n-1}^{-1} C_{2,n-2} C_{4,n-2} \cdots C_{n-4,n-2} & (n:{\rm even}) \\
C_{2,n-1} C_{4,n-1} \cdots C_{n-3,n-1} & (n:{\rm odd}) 
\end{cases}, 
\end{split}
\end{align}
we obtain 
\begin{align}
\begin{split}
\left(\Omega_+ S_+\right)^T \hat{H}_+ \left(\Omega_+ S_+\right) = 
\begin{cases}
\displaystyle
\bigoplus_{r=1}^{\frac{n}{2}} X_{2r-1} & (n:{\rm even}) \\
\displaystyle
\bigoplus_{r=1}^{\frac{n-1}{2}} X_{2r-1} \oplus M_n & (n:{\rm odd}) 
\end{cases}, \\
\left(\Omega_+ S_+\right)^T \hat\Sigma_+ \left(\Omega_+ S_+\right) = 
\begin{cases}
\displaystyle
\bigoplus_{r=1}^{\frac{n}{2}} R_4\left( \frac{2r-1}{n}\pi \right) & (n:{\rm even}) \\
\displaystyle
-i \left( 
\bigoplus_{r=1}^{\frac{n-1}{2}} R_4\left( \frac{2r-1}{n}\pi \right) \oplus (-{\mathbf 1}_2) \right) & (n:{\rm odd}) 
\end{cases}, \\
\end{split}
\end{align}
and 
\begin{align}
\begin{split}
\left(\Omega_- S_-\right)^T \hat{H}_- \left(\Omega_- S_-\right) = 
\begin{cases}
\displaystyle
\bigoplus_{r=1}^{\frac{n-2}{2}} X_{2r}\oplus M_n \oplus M_0 & (n:{\rm even}) \\
\displaystyle
\bigoplus_{r=1}^{\frac{n-1}{2}} X_{2r}\oplus M_0 & (n:{\rm odd}) 
\end{cases}, \\
\left(\Omega_- S_-\right)^T \hat\Sigma_-  \left(\Omega_- S_-\right) = 
\begin{cases}
\displaystyle
-i \left( 
\bigoplus_{r=1}^{\frac{n-2}{2}} R_4\left( \frac{2r}{n}\pi \right)
\oplus (-{\mathbf 1}_2) \oplus {\mathbf 1}_2  \right) & (n:{\rm even}) \\
\displaystyle
\bigoplus_{r=1}^{\frac{n-1}{2}} R_4\left( \frac{2r-1}{n}\pi \right) \oplus {\mathbf 1}_2  & (n:{\rm odd}) 
\end{cases},
\end{split}
\end{align}
where $X_I$ and $R_4(\theta)$ are $4\times 4$ matrices defined by
\begin{equation}
X_I \equiv \left(\begin{matrix} M_I & -N_I \\ N_I & M_I \end{matrix}\right), \quad
R_4(\phi) \equiv R(-\phi) \oplus R(\phi). 
\end{equation}

The matrices $X_I$ and $R_4(\phi)$ are simultaneously transformed into the canonical forms 
as
\begin{equation}
P_I^T X_I P_I = R(-i\epsilon_I \gamma_I) \oplus R(-i\epsilon_I \gamma_I), \quad
P_I^T R_4(\phi) P_I = R(-\phi)\oplus R(\phi), 
\end{equation}
using the $4\times 4$ matrices, 
\begin{equation}
P_I \equiv \left(\begin{matrix}
\cos\frac{\theta_I}{2} & 0 & 0 & \epsilon_I \sin \frac{\theta_I}{2} \\
0 & \cos\frac{\theta_I}{2} & \epsilon_I \sin \frac{\theta_I}{2} & 0 \\
0 & -\epsilon_I \sin\frac{\theta_I}{2} & \cos \frac{\theta_I}{2} & 0 \\
-\epsilon_I \sin\frac{\theta_I}{2} & 0 & 0 & \cos \frac{\theta_I}{2} 
\end{matrix}\right). 
\end{equation}
Thus we define the matrices, 
\begin{equation}
\begin{split}
P_+ \equiv 
\begin{cases}
 \displaystyle
 \bigoplus_{r=1}^{\frac{n}{2}} P_{2r-1} & (n:{\rm even}) \\
 \displaystyle
 \bigoplus_{r=1}^{\frac{n-1}{2}} P_{2r-1} \oplus {\mathbf 1}_2 & (n:{\rm odd}) 
\end{cases}, \quad
P_- \equiv 
\begin{cases}
 \displaystyle
 \bigoplus_{r=1}^{\frac{n-2}{2}} P_{2r} \oplus {\mathbf 1}_4 & (n:{\rm even}) \\
 \displaystyle
 \bigoplus_{r=1}^{\frac{n-1}{2}} P_{2r} \oplus {\mathbf 1}_2 & (n:{\rm odd}) 
\end{cases}. 
\end{split}
\end{equation}

We finally consider the combinations, 
\begin{equation}
T_\pm \equiv \Omega_\pm S_\pm P_\pm, 
\end{equation}
which transform $H_\pm^{m} \Sigma_\pm^{p}$ in the fundamental representation 
into the canonical forms, respectively: \\
\noindent
\fbox{$n$: even}
\begin{align}
\begin{split}
T_+^T &( \hat{H}_+^{m} \hat\Sigma_+^{p} ) T_+ \\ 
&= 
\bigoplus_{r=1}^{\frac{n}{2}} \Bigl( 
R(-im\epsilon_{2r-1} \gamma_{2r-1}-\frac{2r-1}{n}p\pi) \oplus 
R(-im\epsilon_{2r-1} \gamma_{2r-1}+\frac{2r-1}{n}p\pi)
\Bigr), \\
T_-^T &(\hat{H}_-{m} \hat\Sigma_-^{p}) T_- \\
&= 
(-i)^p\biggl\{ \bigoplus_{r=1}^{\frac{n-2}{2}} \Bigl(
R(-im\epsilon_{2r} \gamma_{2r}-\frac{2r}{n}p\pi) \oplus 
R(-im\epsilon_{2r} \gamma_{2r}+\frac{2r}{n}p\pi)
\Bigr) \\
&\hspace{8cm} \oplus R(-im\gamma_n + p\pi) 
\oplus R(\pm im\gamma_0) \biggr\}, 
\end{split}
\label{canonical1}
\end{align}
\noindent
\fbox{$n$: odd}
\begin{align}
\begin{split}
T_+^T &(\hat{H}_+^{m} \hat\Sigma_+^{p}) T_+ \\
&= 
(-i)^p \biggl\{ \bigoplus_{r=1}^{\frac{n-1}{2}} \Bigl(
R(-im\epsilon_{2r-1} \gamma_{2r-1}-\frac{2r-1}{n}p\pi)
\oplus R(-im\epsilon_{2r-1} \gamma_{2r-1}+\frac{2r-1}{n}p\pi)
\Bigr) \\
&\hspace{11cm} 
\oplus R(-im\gamma_n + p\pi) 
\biggr\}, \\
T_-^T &(\hat{H}_-^{m} \hat\Sigma_-^{p}) T_- \\
&= 
\bigoplus_{r=1}^{\frac{n-1}{2}} \Bigl(
R(-im\epsilon_{2r} \gamma_{2r}-\frac{2r}{n}p\pi) \oplus 
R(-im\epsilon_{2r} \gamma_{2r}+\frac{2r}{n}p\pi)
\Bigr)
\oplus R(\pm im\gamma_0), 
\end{split}
\label{canonical2}
\end{align}
where we have used (\ref{M0 and Mn}). 
We can easily see $\det T_\pm=1$. 
These results motivate to introduce (\ref{gamma-c}). 
Note that, when we consider the continuum limit, 
we should approach to the critical 
temperature from the ordered phase. 
Thus the sign appearing in $R(\pm i m \gamma_0)$ is chosen as $+$. 


\section{Partition function of 2D massive fermion on the torus}
\label{app:massive fermion}


Let us consider 2-torus with periods $\omega_1,\omega_2\in {\mathbb C}$ 
and free Majorana fermion with mass $M$ on it: 
\begin{equation}
 S=\frac{1}{2} \int d^2x \Psi^T D \Psi, 
\end{equation}
where $\Psi$ is a two-component spinor, $\Psi$=$\left(\psi,\bar\psi\right)^T$ 
and $D$ is the Dirac matrix given by 
\begin{equation}
 D=\left(\begin{matrix} \del_1+i\del_2 & M \\ -M & \del_1 - i \del_2 \end{matrix} \right). 
\end{equation}
In the following, we evaluate the partition function, 
\begin{equation}
 Z=\int d\Psi e^{-S}={\rm Pf}D, 
 \label{Z massive fermion}
\end{equation}
where ${\rm Pf}D$ denotes the Pfaffian of the Dirac operator $D$.

Let $\{k_1, k_2\}=\{-i\omega_2/A,i\omega_1/A\}$ 
denote the dual vectors corresponding to $\{\omega_1, \omega_2\}$, 
where $A={\rm Im}(\omega_2\bar\omega_1)$ is the area of the torus. 
The plane-wave on the torus is given by 
\begin{align}
 u_{mn}(z,\bar{z})&\equiv 
 e^{2\pi i \left( -n {\rm Re}(\bar{k}_1 z) + m{\rm Re}(\bar{k}_2 z)\right)} \nn \\
 &=e^{\frac{\pi i}{A}\left[ -i(m\bar\omega_1+ n\bar\omega_2)z 
 +i (m\omega_1+n\omega_2)\bar{z} \right]}, 
\end{align}
where $z=x_1+ix_2$ and $\bar{z}=x_1-ix_2$ 
and $m,n$ take integer or half integer,
$m\in \Z + \mu$ and $n \in \Z + \nu$, $(\mu, \nu = 0\ {\rm or}\ 1/2)$, 
depending on the periodicity of the fermion. 
Recall that the eigenvalue of the Laplacian 
$-(\partial_1^2+\partial_2^2)=-4\partial \bar\partial$ 
is 
\begin{align}
 \lambda_{mn}&=\frac{(2\pi)^2}{A^2}\left|m \omega_1 + n \omega_2 \right|^2 \nn \\
 &=
{\mathcal N}^2  \left| m + \tau n \right|^2,
\end{align} 
where ${\mathcal N}=\frac{ \left| 2\pi \omega_1 \right| }{A} $ and 
\begin{equation}
\tau\equiv \omega_2 / \omega_1 = \tau_1 + i \tau_2.  
\quad (\tau_1, \tau_2 \in \R)
\end{equation}
Then the partition function in Eq.(\ref{Z massive fermion}) is expressed as 
\begin{align}
 Z=\sum_{\mu,\nu=0,1/2} Z_{\mu,\nu}, 
\end{align}
with 
\begin{align}
 Z_{\mu,\nu}&=
 \prod_{m,n\in\Z}
{\mathcal N}  \sqrt{ \left| m + n\tau + \mu + \nu \tau \right|^2 + (M/{\mathcal N})^2 }.
 \label{pre Zmn}
\end{align}
Rescaling the mass parameter $M$ as 
\begin{equation}
\mu \equiv \frac{2\pi}{\tau_2} \frac{M}{{\mathcal N}}
\end{equation}
and using the zeta function regularization, 
$Z_{\mu\nu}$ are estimated as as 
\begin{align}
\begin{split}
Z_{0,0}&=
e^{\pi\tau_2 \sum_{n\in\Z}\sqrt{n^2 + \left(\frac{\mu}{2\pi}\right)^2}}
\left(1-e^{-\tau_2 \mu}\right)
\prod_{n=1}^\infty 
\biggl|
1-e^{2\pi i n\left( \tau_1 + i \tau_2 \sqrt{ 1 + \left(\frac{\mu}{2\pi n}\right)^2}\right)}
\biggr|^2,  \\
Z_{0,\frac{1}{2}}&=
e^{\pi\tau_2 \sum_{n\in\Z}\sqrt{\left(n+\frac{1}{2}\right)^2 + \left(\frac{\mu}{2\pi}\right)^2}}
\prod_{n=0}^\infty 
\biggl|
1-e^{\pi i (2\pi+1) \left( \tau_1 + i \tau_2 \sqrt{ 1 + \left(\frac{\mu}{(2n+1)\pi }\right)^2}\right)}
\biggr|^2,  \\
Z_{\frac{1}{2},0}&=
e^{\pi\tau_2 \sum_{n\in\Z}\sqrt{n^2 + \left(\frac{\mu}{2\pi}\right)^2}}
\left(1+e^{-\tau_2 \mu}\right)
\prod_{n=1}^\infty 
\biggl|
1+e^{2\pi i n\left( \tau_1 + i \tau_2 \sqrt{ 1 + \left(\frac{\mu}{2\pi n}\right)^2}\right)}
\biggr|^2,  \\
Z_{\frac{1}{2},\frac{1}{2}}&=
e^{\pi\tau_2 \sum_{n\in\Z}\sqrt{\left(n+\frac{1}{2}\right)^2 + \left(\frac{\mu}{2\pi}\right)^2}}
\prod_{n=0}^\infty 
\biggl|
1+e^{\pi i (2n+1) \left( \tau_1 + i \tau_2 \sqrt{ 1 + \left(\frac{\mu}{(2n+1)\pi }\right)^2}\right)}
\biggr|^2. 
\end{split}
\label{MF Z}
\end{align}
We should note that the exponents of the overall factors in Eq.(\ref{MF Z})
logarithmically diverge;  
\begin{align}
& {\pi\tau_2 \sum_{n\in\Z}\sqrt{n^2 + \left(\frac{\mu}{2\pi}\right)^2}}
 ={-\frac{\pi\tau_2}{6}+\frac{\tau_2 \mu}{2}+\frac{\tau_2 \mu^2}{2\pi}\zeta(1)+\cO(\mu^4)}, \nn \\
 &{\pi\tau_2 \sum_{n\in\Z}\sqrt{\left(n+\frac{1}{2}\right)^2 + \left(\frac{\mu}{2\pi}\right)^2}}
 ={\frac{\pi\tau_2}{12}+\frac{\tau_2 \mu^2}{4\pi}\zeta(1)+\cO(\mu^4)}. 
\end{align}
Thus, precisely speaking, we implicitly regard that  
the logarithmic divergence $\zeta(1)$ is properly regularized 
in Eq.(\ref{MF Z}).


\begin{thebibliography}{99}

\bibitem{Maldacena:1997re} 
  J.~M.~Maldacena,
  ``The Large N limit of superconformal field theories and supergravity,''
  Int.\ J.\ Theor.\ Phys.\  {\bf 38}, 1113 (1999)
  [Adv.\ Theor.\ Math.\ Phys.\  {\bf 2}, 231 (1998)]
  [hep-th/9711200].

\bibitem{Gubser:1998bc} 
  S.~S.~Gubser, I.~R.~Klebanov and A.~M.~Polyakov,
  ``Gauge theory correlators from noncritical string theory,''
  Phys.\ Lett.\ B {\bf 428}, 105 (1998)
  [hep-th/9802109].

\bibitem{Witten:1998qj} 
  E.~Witten,
  ``Anti-de Sitter space and holography,''
  Adv.\ Theor.\ Math.\ Phys.\  {\bf 2}, 253 (1998)
  [hep-th/9802150].

\bibitem{Brown:1986nw} 
  J.~D.~Brown and M.~Henneaux,
  ``Central Charges in the Canonical Realization of Asymptotic 
Symmetries: An Example from Three-Dimensional Gravity,''
  Commun.\ Math.\ Phys.\  {\bf 104}, 207 (1986).

\bibitem{Witten:2007kt} 
  E.~Witten,
  ``Three-Dimensional Gravity Revisited,''
  arXiv:0706.3359 [hep-th].

\bibitem{Maloney:2007ud} 
  A.~Maloney and E.~Witten,
  ``Quantum Gravity Partition Functions in Three Dimensions,''
  JHEP {\bf 1002}, 029 (2010)
  [arXiv:0712.0155 [hep-th]].

\bibitem{Gaberdiel:2010pz} 
  M.~R.~Gaberdiel and R.~Gopakumar,
  ``An AdS$_3$ Dual for Minimal Model CFTs,''
  Phys.\ Rev.\ D {\bf 83}, 066007 (2011)
  [arXiv:1011.2986 [hep-th]].

\bibitem{Gaberdiel:2012uj} 
  M.~R.~Gaberdiel and R.~Gopakumar,
  ``Minimal Model Holography,''
  J.\ Phys.\ A {\bf 46}, 214002 (2013)
  [arXiv:1207.6697 [hep-th]].

\bibitem{Henneaux:2010xg} 
  M.~Henneaux and S.~J.~Rey,
  ``Nonlinear $W_{infinity}$ as Asymptotic Symmetry of Three-Dimensional Higher Spin Anti-de Sitter Gravity,''
  JHEP {\bf 1012}, 007 (2010)
  [arXiv:1008.4579 [hep-th]].

\bibitem{Campoleoni:2010zq} 
  A.~Campoleoni, S.~Fredenhagen, S.~Pfenninger and S.~Theisen,
  ``Asymptotic symmetries of three-dimensional gravity coupled to higher-spin fields,''
  JHEP {\bf 1011}, 007 (2010)
  [arXiv:1008.4744 [hep-th]].

\bibitem{Gaberdiel:2011wb} 
  M.~R.~Gaberdiel and T.~Hartman,
  ``Symmetries of Holographic Minimal Models,''
  JHEP {\bf 1105}, 031 (2011)
  [arXiv:1101.2910 [hep-th]].


\bibitem{Campoleoni:2011hg} 
  A.~Campoleoni, S.~Fredenhagen and S.~Pfenninger,
  ``Asymptotic W-symmetries in three-dimensional higher-spin gauge theories,''
  JHEP {\bf 1109}, 113 (2011)
  [arXiv:1107.0290 [hep-th]].

\bibitem{Creutzig:2011fe} 
  T.~Creutzig, Y.~Hikida and P.~B.~Ronne,
  ``Higher spin AdS$_3$ supergravity and its dual CFT,''
  JHEP {\bf 1202}, 109 (2012)
  [arXiv:1111.2139 [hep-th]].

\bibitem{Castro:2010ce} 
  A.~Castro, A.~Lepage-Jutier and A.~Maloney,
  ``Higher Spin Theories in AdS$_3$ and a Gravitational Exclusion Principle,''
  JHEP {\bf 1101}, 142 (2011)
  [arXiv:1012.0598 [hep-th]].

\bibitem{Gaberdiel:2012yb} 
  M.~R.~Gaberdiel, T.~Hartman and K.~Jin,
  ``Higher Spin Black Holes from CFT,''
  JHEP {\bf 1204}, 103 (2012)
  [arXiv:1203.0015 [hep-th]].

\bibitem{Gaberdiel:2012ku} 
  M.~R.~Gaberdiel and R.~Gopakumar,
  ``Triality in Minimal Model Holography,''
  JHEP {\bf 1207}, 127 (2012)
  [arXiv:1205.2472 [hep-th]].

\bibitem{Perlmutter:2012ds} 
  E.~Perlmutter, T.~Prochazka and J.~Raeymaekers,
  ``The semiclassical limit of W$_N$ CFTs and Vasiliev theory,''
  JHEP {\bf 1305}, 007 (2013)
  [arXiv:1210.8452 [hep-th]].

\bibitem{Chang:2013izp} 
  C.~M.~Chang and X.~Yin,
  ``A semilocal holographic minimal model,''
  Phys.\ Rev.\ D {\bf 88}, no. 10, 106002 (2013)
  [arXiv:1302.4420 [hep-th]].

\bibitem{Ferlaino:2013vga} 
  M.~Ferlaino, T.~Hollowood and S.~P.~Kumar,
  ``Asymptotic symmetries and thermodynamics of higher spin 
black holes in AdS3,''
  Phys.\ Rev.\ D {\bf 88}, 066010 (2013)
  [arXiv:1305.2011 [hep-th]].

\bibitem{Gaberdiel:2013jpa} 
  M.~R.~Gaberdiel, K.~Jin and W.~Li,
  ``Perturbations of W(infinity) CFTs,''
  JHEP {\bf 1310}, 162 (2013)
  [arXiv:1307.4087].

\bibitem{Fujisawa:2013ima} 
  I.~Fujisawa, K.~Nakagawa and R.~Nakayama,
  ``AdS/CFT for 3D Higher-Spin Gravity Coupled to Matter Fields,''
  Class.\ Quant.\ Grav.\  {\bf 31}, 065006 (2014)
  [arXiv:1311.4714 [hep-th]].

\bibitem{Creutzig:2014ula} 
  T.~Creutzig, Y.~Hikida and P.~B.~Ronne,
  ``Higher spin AdS$_{3}$ holography with extended supersymmetry,''
  JHEP {\bf 1410}, 163 (2014)
  [arXiv:1406.1521 [hep-th]].


\bibitem{Banks:1996vh} 
  T.~Banks, W.~Fischler, S.~H.~Shenker and L.~Susskind,
  ``M theory as a matrix model: A Conjecture,''
  Phys.\ Rev.\ D {\bf 55}, 5112 (1997)
  [hep-th/9610043].

\bibitem{Hanada:2008ez} 
  M.~Hanada, Y.~Hyakutake, J.~Nishimura and S.~Takeuchi,
  ``Higher derivative corrections to black hole thermodynamics 
from supersymmetric matrix quantum mechanics,''
  Phys.\ Rev.\ Lett.\  {\bf 102}, 191602 (2009)
  [arXiv:0811.3102 [hep-th]].

\bibitem{Hanada:2013rga} 
  M.~Hanada, Y.~Hyakutake, G.~Ishiki and J.~Nishimura,
  ``Holographic description of quantum black hole on a computer,''
  Science {\bf 344}, 882 (2014)
  [arXiv:1311.5607 [hep-th]].

\bibitem{Hyakutake:2007sm} 
  Y.~Hyakutake,
  ``Toward the Determination of R**3 F**2 Terms in M-theory,''
  Prog.\ Theor.\ Phys.\  {\bf 118}, 109 (2007)
  [hep-th/0703154 [HEP-TH]].

\bibitem{Castro:2011zq} 
  A.~Castro, M.~R.~Gaberdiel, T.~Hartman, A.~Maloney and R.~Volpato,
  ``The Gravity Dual of the Ising Model,''
  Phys.\ Rev.\ D {\bf 85}, 024032 (2012)
  [arXiv:1111.1987 [hep-th]].
  
\bibitem{Akhmedov:1998vf} 
  E.~T.~Akhmedov,
  ``A Remark on the AdS / CFT correspondence and the 
renormalization group flow,''
  Phys.\ Lett.\ B {\bf 442}, 152 (1998)
  [hep-th/9806217].
  
\bibitem{Alvarez:1998wr} 
  E.~Alvarez and C.~Gomez,
  ``Geometric holography, the renormalization group and the c theorem,''
  Nucl.\ Phys.\ B {\bf 541}, 441 (1999)
  [hep-th/9807226].
  
\bibitem{Freedman:1999gp} 
  D.~Z.~Freedman, S.~S.~Gubser, K.~Pilch and N.~P.~Warner,
  ``Renormalization group flows from holography supersymmetry 
and a c theorem,''
  Adv.\ Theor.\ Math.\ Phys.\  {\bf 3}, 363 (1999)
  [hep-th/9904017].
  
\bibitem{Girardello:1998pd} 
  L.~Girardello, M.~Petrini, M.~Porrati and A.~Zaffaroni,
  ``Novel local CFT and exact results on perturbations of N=4 
superYang Mills from AdS dynamics,''
  JHEP {\bf 9812}, 022 (1998)
  [hep-th/9810126].
  
\bibitem{Girardello:1999bd} 
  L.~Girardello, M.~Petrini, M.~Porrati and A.~Zaffaroni,
  ``The Supergravity dual of N=1 superYang-Mills theory,''
  Nucl.\ Phys.\ B {\bf 569}, 451 (2000)
  [hep-th/9909047].
  
\bibitem{Porrati:1999ew} 
  M.~Porrati and A.~Starinets,
  ``RG fixed points in supergravity duals of 4-D field theory 
and asymptotically AdS spaces,''
  Phys.\ Lett.\ B {\bf 454}, 77 (1999)
  [hep-th/9903085].
  
\bibitem{Balasubramanian:1999jd} 
  V.~Balasubramanian and P.~Kraus,
  ``Space-time and the holographic renormalization group,''
  Phys.\ Rev.\ Lett.\  {\bf 83}, 3605 (1999)
  [hep-th/9903190].
  
\bibitem{Skenderis:1999mm} 
  K.~Skenderis and P.~K.~Townsend,
  ``Gravitational stability and renormalization group flow,''
  Phys.\ Lett.\ B {\bf 468}, 46 (1999)
  [hep-th/9909070].
  
  
\bibitem{DeWolfe:1999cp} 
  O.~DeWolfe, D.~Z.~Freedman, S.~S.~Gubser and A.~Karch,
  ``Modeling the fifth-dimension with scalars and gravity,''
  Phys.\ Rev.\ D {\bf 62}, 046008 (2000)
  [hep-th/9909134].
  
\bibitem{de Boer:1999xf} 
  J.~de Boer, E.~P.~Verlinde and H.~L.~Verlinde,
  ``On the holographic renormalization group,''
  JHEP {\bf 0008}, 003 (2000)
  [hep-th/9912012].
  
  
\bibitem{Fukuma:2000bz} 
  M.~Fukuma, S.~Matsuura and T.~Sakai,
  ``A Note on the Weyl anomaly in the holographic renormalization group,''
  Prog.\ Theor.\ Phys.\  {\bf 104}, 1089 (2000)
  [hep-th/0007062].

\bibitem{Fukuma:2002sb} 
  M.~Fukuma, S.~Matsuura and T.~Sakai,
  ``Holographic renormalization group,''
  Prog.\ Theor.\ Phys.\  {\bf 109}, 489 (2003)
  [hep-th/0212314].


\bibitem{onsager} 
  L. Onsager, 
``Crystal Statistics. I. A Two-Dimensional Model with an 
Order-Disorder Transition,'' Phys.\ Rev.\ {\bf 65}, 117 (1944).

\bibitem{kaufman} 
  B. Kaufman, 
``Crystal Statistics. II. Partition Function Evaluated 
by Spinor Analysis,'' Phys.\ Rev.\ {\bf 76}, 1232 (1949).

\bibitem{ferdinandFisher} 
  A.~E.~Ferdinand and M.~E.~Fisher, 
``Bounded and Inhomogeneous Ising Models. I. Specific-Heat 
Anomaly of a Finite Lattice,'' Phys.\ Rev.\ {\bf 185}, 832 (1969).

\bibitem{Melzer:1993zk} 
  E.~Melzer,
  ``Fermionic character sums and the corner transfer matrix,''
  Int.\ J.\ Mod.\ Phys.\ A {\bf 9}, 1115 (1994)
  [hep-th/9305114].

\bibitem{O'Brien:1996} 
  D.~L.~O'Brien, P.~A.~Pearce, and S.~O.~Warnaar, 
  ``Finitized conformal spectrum of the Ising model o the 
cylinder and torus,''
  Physica A {\bf 228}, 63 (1996)


\bibitem{Chui:2001nu} 
  C.~H.~O.~Chui and P.~A.~Pearce,
  ``Finitized conformal spectra of the Ising model on the 
Klein bottle and Mobius strip,''
  J.\ Statist.\ Phys.\  {\bf 107}, 1167 (2002)
  [hep-th/0105233].

\bibitem{Feverati:2004bv} 
  G.~Feverati and P.~Grinza,
  ``Integrals of motion from TBA and lattice-conformal dictionary,''
  Nucl.\ Phys.\ B {\bf 702}, 495 (2004)
  [hep-th/0405110].

\bibitem{Hotta:2008xt} 
  K.~Hotta, Y.~Hyakutake, T.~Kubota, T.~Nishinaka and H.~Tanida,
  ``The CFT-interpolating Black Hole in Three Dimensions,''
  JHEP {\bf 0901}, 010 (2009)
  [arXiv:0811.0910 [hep-th]].
  
\bibitem{Codello:2013iqa} 
  A.~Codello, G.~D'Odorico and C.~Pagani,
  ``A functional RG equation for the c-function,''
  JHEP {\bf 1407}, 040 (2014)
  [arXiv:1312.7097 [hep-th]].

  
  
\bibitem{Breitenlohner:1982jf} 
  P.~Breitenlohner and D.~Z.~Freedman,
  ``Stability in Gauged Extended Supergravity,''
  Annals Phys.\  {\bf 144}, 249 (1982).

\bibitem{Zamolodchikov:1986gt} 
  A.~B.~Zamolodchikov,
  ``Irreversibility of the Flux of the Renormalization Group 
in a 2D Field Theory,''
  JETP Lett.\  {\bf 43}, 730 (1986)
  [Pisma Zh.\ Eksp.\ Teor.\ Fiz.\  {\bf 43}, 565 (1986)].


\bibitem{nernst} S.A. Hartnoll, P.K. Kovtun, M. Mueller, S. Sachdev, 
``Theory of the Nernst effect near quantum phase transitions 
in condensed matter, and in dyonic black holes,'' 
Phys.Rev. {\bf B76} 144502 (2007) [arXiv:0706.3215 [cond-mat]]. 


\bibitem{Hartnoll:2012rj} 
  S.~A.~Hartnoll and D.~M.~Hofman,
  ``Locally Critical Resistivities from Umklapp Scattering,''
  Phys.\ Rev.\ Lett.\  {\bf 108}, 241601 (2012)
  [arXiv:1201.3917 [hep-th]].


\bibitem{Blake:2013owa} 
  M.~Blake, D.~Tong and D.~Vegh,
  ``Holographic Lattices Give the Graviton an Effective Mass,''
  Phys.\ Rev.\ Lett.\  {\bf 112}, no. 7, 071602 (2014)
  [arXiv:1310.3832 [hep-th]].

\bibitem{Blake:2014yla} 
  M.~Blake and A.~Donos,
  ``Quantum Critical Transport and the Hall Angle,''
  Phys.\ Rev.\ Lett.\  {\bf 114}, no. 2, 021601 (2015)
  [arXiv:1406.1659 [hep-th]].

\bibitem{Baggioli:2014roa} 
  M.~Baggioli and O.~Pujolas,
  ``Electron-Phonon Interactions, Metal-Insulator Transitions, 
and Holographic Massive Gravity,''
  Phys.\ Rev.\ Lett.\  {\bf 114}, no. 25, 251602 (2015)
  [arXiv:1411.1003 [hep-th]].

 \end{thebibliography}
\end{document}